# Pressure-Thresholded Response in Cylindrically Shocked Cyclotrimethylene Trinitramine (RDX)

*Leora E. Dresselhaus-Cooper,* [\*ab†] *Dmitro Martynowych,* [ab] *Fan Zhang,* [d] *Charlene Tsay,* [e] *Jan Ilavsky,* [f] *SuYin Grass Wang,* [g] *Yu-Sheng Chen,* [g] *Keith A. Nelson* [ab]

[a] Department of Chemistry, Massachusetts Institute of Technology, 77 Massachusetts Ave., Cambridge, MA 02139

[b] Institute for Soldier Nanotechnology, Massachusetts Institute of Technology, 77 Massachusetts Ave., Cambridge, MA 02139

[†] Present address: [c] Physics Division, Lawrence Livermore National Laboratory, 7000 East Ave. L-487, Livermore, CA 94550

[d] Materials Measurement Science Division, National Institute of Standards and Technology, 100 Bureau Dr., Gaithersburg, MD 20899

[e] Department of Chemistry, University of California Riverside, 501 Big Springs Rd., Riverside, CA 92521

[f] X-ray Science Division, Argonne National Laboratory, 9700 S. Cass Ave., Argonne, IL 60439

[g] ChemMatCARS, Center for Advanced Radiation Sources, The University of Chicago, 9700 S. Cass Ave., Argonne, IL 60439

ABSTRACT.

We demonstrate a strongly thresholded response in cyclotrimethylene trinitramine (RDX) when it is cylindrically shocked using a novel waveguide geometry. Using ultrafast single-shot multi-frame imaging, we demonstrate that <100-μm diameter single crystals of RDX embedded in a polymer host deform along preferential planes for >100 ns after the shock first arrives in the crystal. We use *in-situ* imaging and time-resolved photoemission to demonstrate that short-lived chemistry occurs with complex deformation pathways. Using scanning electron microscopy and ultrasmall-angle X-ray scattering, we demonstrate that the shock-induced dynamics leave behind porous crystals, with pore shapes and sizes that change significantly with shock energy. A threshold pressure of ~ 12 GPa at the center of convergence separated the single-mode planar crystal deformations from the chemistry-coupled multi-plane dynamics at higher pressures. Our observations indicate preferential directions for deformation for our cylindrically shocked system, despite the applied stress along many different crystallographic planes.

INTRODUCTION

Energetic materials are a class of reactive compounds that have powerful decomposition pathways that can be activated by mechanical drivers[1]. Cyclotrimethylene trinitramine (RDX) is one such energetic material, which has been used extensively for engineering, construction and armament applications since it was developed in World War I[2]. In practical applications, RDX crystals are packed into a polymer binder to form *polymer bonded explosives (PBX)* whose kinetics and detonation powers depend on their compositions. However, the mechanistic link between chemistry and mechanics has been elusive[3]. The hot-spot model, first developed by Bowden and Yoffe, predicts that the chemistry in energetic materials initiates at locally high-temperature "hotspots".[4] Tarver, *et al.* found the critical conditions necessary for hotspots to reach self-sustaining levels (i.e. denotation) at the macroscale [5]. However, many uncertainties still exist about the microscopic mechanisms that activate critical or sub-critical hotspots [9–13].

Early work on RDX demonstrated that explosives with more initial defects are more sensitive to shock-induced chemistry[14,15]. Ductile[16,17] and brittle[18,19] responses in RDX have been predicted and observed under different conditions. Different chemical mechanisms have been observed for decomposition activated at different strain rates[20–22]. Shock waves produce the highest attainable strain rates of mechanically-driven systems, and the extreme pressure-temperature states that they create can induce exotic chemistry that is not accessible at ambient conditions[23]. Since shock waves irreversibly destroy materials through pathways that are extremely sensitive to variation among samples down to the nanoscale, shock responses of materials are notoriously difficult to characterize[24]. Detailed measurements of shock-induced reactive pathways have been limited, despite the practical needs for the scientific insight that such measurements could provide.

Molecular dynamics (MD) studies by Cawkwell, *et al.*, have demonstrated that shocks in RDX along the <001> axis form shear bands, as their slip systems cannot accommodate the strain[16,25]. Theoretical work by other groups has also indicated pressure-dependent anisotropic plasticity in uniaxially shocked RDX[26–29]. Near-equilibrium indentation tests have probed the active deformation planes in RDX, revealing fourteen active planes in the *Pbca* crystal (*a* = 13.182 Å, *b* = 11.574 Å, *c* = 10.709 Å)[30], with localized plasticity and fracture around the indentation tip[17,31–33]. At large length scales, fracto-emission has been observed as detonations are initiated in RDX[18], while shocked, small RDX crystals have shown additional slip, cleavage and shear bands after shock loading[34,35]. Several different chemical decomposition mechanisms are known to compete, and predictions have shown different relative extents based on the shock pressure[36,37]. While chemistry and plastic transformations have always been predicted and observed to occur together, a direct link between the detailed plastic modes and the chemical pathways has remained elusive.

Our view of the dynamics induced by idealized uniaxial shocks is still incomplete, but real-world PBXs experience more complex, highly non-planar shock geometries. The heterogeneity in a PBX creates many interfaces, resulting in multiple reflections and refractions that distort the planarity of an initially planar shock wave[12]. While experiments on simple materials have explored how the geometry of a shock wave influences the material dynamics that ensue[38,39], the effects of non-planar shocks on energetic materials remain largely unknown. Work by An, *et al.*[12,40] has modeled how heterogeneities in PBX materials cause an initially planar shock to deform into non-planar waves, which initiate hotspots to different extents than the planar waves. To understand how RDX decomposes in PBXs and to inform the design of mechanically driven chemical systems, detailed experimental studies must be extended to non-planar shock waves.

This work examines how cylindrically converging shock waves initiate deformations and chemistry in RDX. We use a novel quasi-2D waveguide shock geometry that enables us to observe how RDX crystals that are immobilized in a polystyrene matrix respond to cylindrical shocks. With single-shot multi-frame image sequences, we track the progression of the deformations induced by converging shock waves. We see a high degree of crystallographic alignment in the deformations, despite the many directions along which the shock stresses are exerted. Scanning electron microscopy (SEM) and ultra-small angle X-ray scattering (USAXS) demonstrate that the crystals recovered after they are shocked have pore structures with sizes and morphologies that depend strongly on the shock pressure. Real-time imaging and photoemission results indicate a strong pressure threshold in the shock-induced chemistry and deformation. Above a ~12 GPa pressure threshold, we observe a cascade of deformation planes which appear and shift for >20 ns after the shock has passed and which are correlated with increased chemical decomposition. Our *in-situ* imaging and fluorescence results as well as our *ex-situ* X-ray scattering results provide new insights into the links between mechanical deformation and chemistry in shocked RDX.

## EXPERIMENTAL METHODS

### I.    *Shock Experimentation and Sample Preparation*

All waveguide shocks in this work were produced by pulsed laser excitation (150 ps, 1-9 mJ, 800 nm) with a Gaussian drive laser beam that was passed through a 0.5º axicon and an $f = 3$ mm lens to form a ring-shaped drive with 150 μm diameter at the sample.[41] An absorbing laser dye (Epolin 3036[i]) dissolved in the polystyrene sample absorbed the drive laser light, initiating thermal expansion that launched the compressive wave. Real-time images of the shock-induced dynamics were collected with single-shot multi-frame imaging, with femtosecond or nanosecond integration times to probe different timescales (Fig. 1a). For femtosecond multi-frame imaging, (Fig. 1b, bottom) the compressed 800 nm, 130 fs laser output was passed through a frequency-doubling Fabry-Perot cavity to create a train of 400-nm femtosecond imaging probe pulses, spaced 3-5 ns apart in time.[42] Nanosecond multi-frame image sequences (Fig. 1b, top) were collected with a 10-μs duration, 670-nm wavelength SILUX diode laser as the light source, using the camera to electronically gate the pulse. Both imaging configurations (Fig. 1a) included an ultrahigh-frame-rate SIM 16X camera to collect the image sequences in real time. A single-frame Hamamatsu Orca camera was used to align the crystals and image the recovered samples.

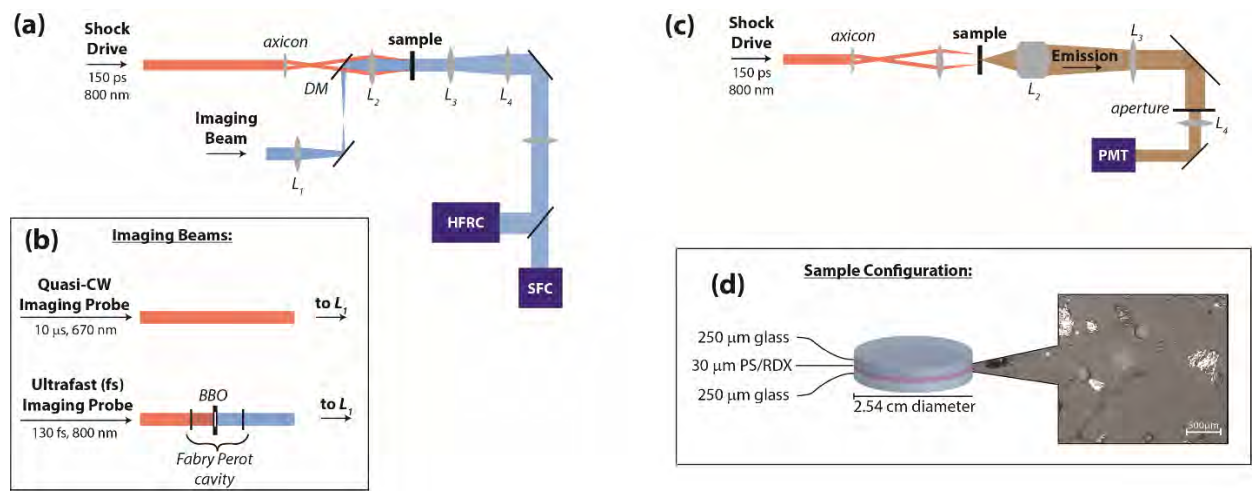

**Figure 1.** (a) Schematic illustration of the shock and imaging configurations used to collect real-time images with a high frame-rate camera (HFRC), and the "before" and "after" frames on a single-frame camera (SFC); (b) Schematic illustrations of the uncollimated quasi-CW and femtosecond imaging beams before reaching lens L1. (c) Optical configuration used to measure shock-induced emission. (d) The sample geometry, with an image of part of a typical sample.

Shock-induced emission from the substrates, the polymer and the RDX were measured in a similar configuration, as shown in Fig. 1c. The imaging probe beam was blocked, and the light emitted from the sample was imaged with a 10X objective and a subsequent $f$ = 1000 mm lens onto an aperture. The aperture spatially filtered the emitted light, allowing us to remove most of the laser-induced fluorescence from the excitation "ring" region (some persisted, due to scatter at the aperture). Light passing through aperture was imaged with an $f = 100$ mm lens onto a photomultiplier tube and the resulting signal was collected and saved on a 4 GHz bandwidth oscilloscope. For each drive pulse energy, emission profiles were measured from the shocked polymer with and without an RDX crystal inside the shock ring, with repeated measurements at each drive pulse energy to reveal reproducible features of the responses.

Samples were made by embedding a collection of small (25 μm to 300 μm dimensions) RDX crystals into a matrix of polystyrene with a dissolved 800 nm absorptive laser dye (Epolin 3036). The small crystals were grown under ambient conditions by slow evaporation from a solvent mixture made by adding dimethyl formamide (18% vol.) to a saturated solution of RDX in acetonitrile. The resulting crystals were suspended in a solution of polystyrene (15% mass) and Epolin 3036 (1% mass) in toluene (84% mass) and were drop-cast onto 2.54 cm diameter fused silica microscope slides (250 μm thick). Polystyrene was chosen for the polymer matrix because it had good adhesion to the RDX crystals, and its acoustic impedance nearly matched that of RDX. After drying for 48 h, the sample layer was polished to a ~ 50 μm thickness using aluminum oxide lapping paper in a water-alkonox solution with a minimum grit size of 500 nm. Samples with the "capped" geometry were made by placing a thin layer of glycerol over the polished film (the capping fluid) and another 250 μm thick fused silica slide to seal the sample. Samples of the "uncapped" geometry were made by rewetting the polished film with a thin layer of toluene and pressing the second fused silica substrate atop the damp sample layer to create a tight seal. After they were sealed, the uncapped samples were fitted inside a compressive clamp and dried under vacuum for 12 h. While the capped samples had an accessible surface for some *ex-situ* characterization methods that required removal of the top substrate, the low impedance of the capping fluid reduced the confinement of the shocks. Uncapped samples had good surface contact between the sample and both substrates, providing high optical quality and shock confinement in the waveguide geometry. The thickness of each uncapped sample was measured with a Keyence CK-X200 confocal microscope. Specific sample thicknesses and shock pressures were 37.5 μm (7 GPa), 28 μm (10 GPa), 34 μm (16 GPa) and 33 μm (28 GPa).

RDX crystals in the polystyrene matrix had slight mobility as they were polished, which created poor surface quality in crystals that were located at the polished surface of the sample layer. To ensure that our measured RDX responses were not influenced by mechanical perturbation and crystal surface damage during polishing, we limited all experiments (except for SEM measurements) to embedded crystals that were never in direct contact with the lapping paper. A single sample assembly typically contained 50-500 RDX crystals that were suitable for shock measurements.

The highly heterogeneous samples were placed in the "Sample" position in the optical configurations shown in Fig. 1 and were then translated using a three-dimensional sample stage. We used the single-frame camera to locate RDX crystals with ≤ 125 μm diameter and to position them to receive a converging shock launched in the surrounding polymer. We then used a shutter and a set of Stanford Research Systems DG535 timing boxes to synchronize our drive beam and multi-frame camera, and we collected image sequences of the shock travelling through the polymer and the RDX crystal as well as subsequent shock-induced responses.

II.     *Determination of Pressure*

The shocks within the RDX crystals did not produce discernable imaging signals in the uncapped geometry (likely due to small photoelastic constants, as described in the Results section). However, we could monitor shock propagation in the polymer matrix before the shock entered a crystal, and from the shock velocity in the polymer we could estimate the shock velocity ($U_s$) and pressure ($P$) states reached in the RDX crystals in uncapped samples. We used a boundary-detection image processing algorithm called locally adaptive discriminant analysis (LADA)[43,44] to locate the shock in each frame while it was still in the polymer. Knowing the time between frames, we could determine the average velocity of the shock wave, $U_s$, as it converged toward an RDX crystal. The impedance-matching method allowed us to convert the measured polymer $U_s$ value to a corresponding RDX $U_s$ value at the interface, assuming the shocked states lie along the principal Hugoniot of each material.[45] We used the literature parameters for the shock Hugoniot of RDX, but adjusted the acoustic velocity of polystyrene to the $C_0 = 1.1$ km/s value that we measured for the waveguide system.

**Table 1.** Predicted shock velocities and pressures, assuming transduction at $R_s = 20$ μm, shocked states using the principal Hugoniot, and the CCW model for acceleration.

| Drive Laser Energy [mJ] | $U_s$ in Polystyrene at Interface with RDX (measured by LADA) [km/s] | $U_s$ in RDX at Interface ($R_s = 20$ μm) [km/s] | Interface Pressure of RDX on Principal Hugoniot [GPa] | $U_s$ in RDX at $R_s = 5$ μm (10% unc. in $U_s$ before CCW) [km/s] | Pressure at $R_s = 5$ μm on Principal Hugoniot (10% unc. in $U_s$ before CCW) [GPa] |
|---|---|---|---|---|---|
| 1.5 | 1.6 | 3.09 ± 0.02 | 0.92 ± 0.085 | 4.5 ± 0.2 | 7 ± 3 |
| 2.5 | 2.0 | 3.41 ± 0.04 | 2.0 ± 0.15 | 4.9 ± 0.3 | 10 ± 3 |
| 3.5 | 2.3 | 3.67 ± 0.04 | 3.1 ± 0.2 | 5.7 ± 0.3 | 16 ± 5 |
| 4.5 | 2.5 | 3.8 ± 0.25 | 3.9 ± 0.7 | 7.0 ± 0.4 | 28 ± 7 |

Using the calculated $U_s$ value in RDX at the interface as the initial condition, we used the Chester-Chisnell-Whitham (CCW) model[46–48] of the accelerating shock to calculate the $U_s$ values in RDX at all times as the shock converged.[48] While the model yields shock velocity values that do not necessarily lie along the principal Hugoniot, we used the Hugoniot to relate our predicted shock velocities to the corresponding shock pressures at the shock front. Our previous work[42] showed that the converging geometry and edge-effects from partial confinement in the waveguide structure cause the actual shock pressure to deviate by 2-10% from the Hugoniot as the shock approaches the center of convergence for 30 μm thick waveguides.[42] We also estimated a shock focal spot size of 5 μm diameter. This value is based on previous studies which demonstrated that our converging waveguide shocks deviate substantially from the principal Hugoniot for $R_s < 5$ μm.[42] A strong deviation from the Hugoniot is inevitable in a focusing shock, as the pressure is highest at the focus but the net particle velocity is zero. The estimated values for the shock parameters resulting from different shock drive laser pulse energies, and the uncertainties in the shock parameters, are shown in Table 1. Some reassurance that the uncertainties are not greater than those indicated in the table comes from the fact that as presented below, we observed different pressure-dependent responses repeatedly for different RDX crystals, depending with high consistency on the drive laser pulse energies. We also observed similar responses extending across significant regions of shocked crystals, clearly indicating that there is not an extremely sharp focus in one region at which the pressure is highest. Altogether we conducted observations on 5-10 different RDX crystals for each of the four drive pulse energies listed below.

RESULTS

## I.    Ultrashort-Window Images of Shock Propagation and Response

The morphology and velocity of the converging shock in the RDX crystals were most clearly observable in image sequences of capped samples, as shown in Fig. 2. The blue and red lines at 18 ns and 23 ns show the angular components of the converging shock ring that traverse the RDX crystal and polystyrene, respectively (see Supplement for additional details). Comparison between the red and blue lines in Fig. 2 reveals that the shock velocity is highest for the components that traverse the stiffer RDX crystal than the surrounding polymer. As the crystal is non-cylindrical, the shape of the shock ring changes as it traverses the RDX, with leading angular components corresponding to the components of the shock that have traversed the longest distance in the crystal. The change in curvature and velocity of the shock wave verifies that the shock wave transmits (at least in part) into the RDX crystals.

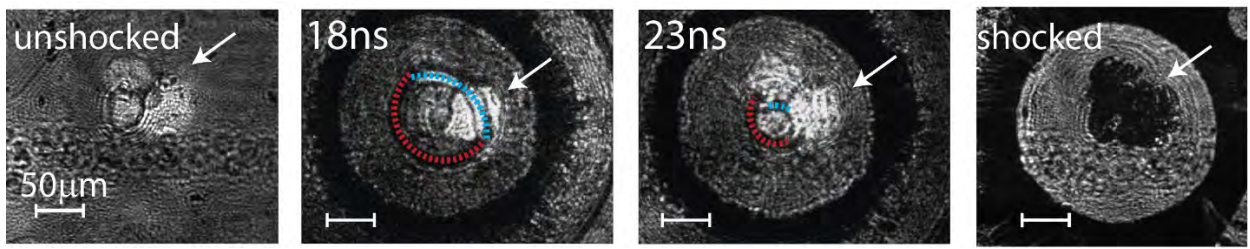

**Figure 2.** Direct shadowgraph images showing a shock wave produced by drive laser pulse energy $E_{drive} = 3.5$ mJ in a capped sample as it traverses an RDX crystal. The crystal is indicated by the white arrow, the shock in polystyrene is shown with red dashed lines, and the shock in RDX is shown with blue dashed lines. The images were collected using femtosecond-duration probe pulses (as shown in the lower part of Fig. 1b) to resolve the propagating shock wave without blurring. Images were collected without crossed polarizers.

The unambiguous change in velocity and curvature in Fig. 2 are only observable in capped samples, as the shock in RDX is coupled to the capping fluid whose high photoelastic constant produces a clear feature in the image. The capping fluid reduces the degree of confinement of the shock within the RDX sample layer, modifying the material dynamics. While the capped samples were necessary to directly validate that the shock transmits into RDX crystals, the uncapped samples were required for a better constrained P-$U_s$ relationship (Section IV, Supporting Information). All of our analysis of shock-induced RDX responses comes from uncapped samples.

An image sequence of a similar converging shock traversing an RDX crystal in the uncapped geometry is shown in Fig. 3, revealing the material dynamics following the shock. The shock wave is evident in frames collected at $t$ = 38 ns, 41 ns, and 44 ns, as the set of concentric rings inside the laser excitation ring. At $t$ = 38 ns the wave is converging, with the shock front corresponding to the innermost ring. Contrast in the direct shadowgraph images is described by the Laplacian of the sample's refractive index, showing a shock as a bright then dark intensity feature which scales with the density variation, $\frac{\partial^2 \rho}{\partial r^2}$, at the shock front and release, respectively. As the shock front appears as the bright leading edge of the shock feature, it is evident in Fig. 3 that the wave is diverging at $t$ = 41 ns and 44 ns.

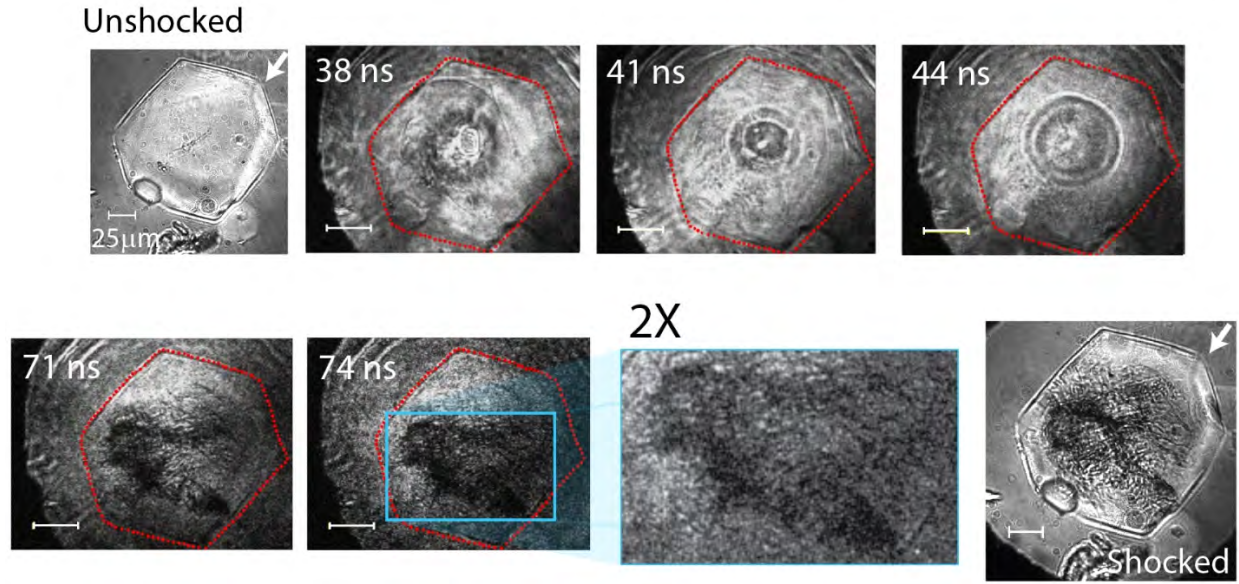

**Figure 3.** Direct shadowgraph images of an ~ 7 GPa shock produced by $E_{\text{drive}}$ = 2.0 mJ traversing a crystal of RDX in the uncapped sample geometry. The white arrow in each frame indicates the RDX crystal, which is outlined in red. An expanded and brightened view of the damaged region of the crystal in the frame at $t$ = 74 ns. Images were collected without crossed polarizers.

Close examination of the initial three frames in Fig. 3 shows that the RDX crystal in these frames contains a large population of poorly resolved, somewhat indiscernible shapes. Indistinct image features like those seen in Fig. 3 can correspond either to objects that are significantly out of the focal plane of the imaging system or to objects that are slightly smaller than the resolution of the imaging configuration. Further imaging of the recovered crystal with confocal scanning in a diffraction-limited microscope (magnification ≤ 100x) demonstrated that these features cannot be resolved for any focal plane in the sample. Given the 0.28 numerical aperture (NA) for the *in-situ* imaging configuration in Fig. 3, the diffraction limit with our 400-nm illumination was $\frac{\lambda}{2*NA}$ = 710 nm, placing an upper bound on the size of the features. As is evident at 38 ns and 41 ns, the sub-micrometer features appear radially a distance ~ 4 μm after the leading edge of the shock front. By dividing the radial distance between the shock front and the initial RDX response by the average shock velocity (~4 km/s), we see that the sub-micrometer features appear within the first 1 ns of the shock front. Given the ~15-20 ns duration of the elevated pressure in the waveguide converging shocks[49], the results in Fig. 3 indicate that sub-micrometer features appear before the material returns to ambient pressure.

*II.    Analysis of Crystals Recovered after Shock*

To investigate the shock-induced nanoscale features produced in the RDX, we used scanning electron microscopy (SEM) and ultra-small angle X-ray scattering (USAXS) on recovered crystals. Together, these two measurements reveal much about the morphologies and the statistics of the void structures formed by the shock waves.

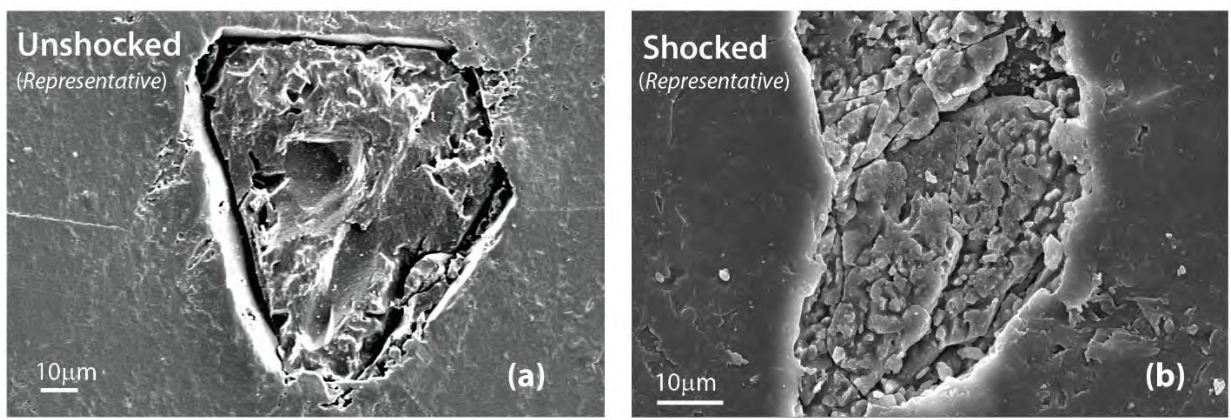

**Figure 4.** SEM images of two representative RDX crystals, (a) one before and (b) the other after shocking. The recovered crystal clearly indicates the porous structure seen after shocking. The drive laser pulse energy was 3.5 mJ for an RDX crystal embedded in a capped sample.

SEM analysis was conducted on shocked crystals in capped samples because these had accessible polished crystal faces at the sample surfaces. Crystal motion in the polymer host as the samples were polished worsened the surface smoothness, but representative populations of shocked and unshocked RDX crystals revealed distinct differences in their SEM images. A representative unshocked crystal can be seen in Fig. 4a, while a representative crystal recovered after shock is shown in Fig. 4b.

Crystal facets are evident in the unshocked crystal in Fig. 4a, indicating its single crystallinity (as confirmed by X-ray diffraction). The morphology of the shocked crystal in Fig. 4b indicates an intricate network of cracks and voids covering length scales from micrometers to nanometers. Similar porous structures have been observed in RDX crystals upon slow thermal decomposition[50] because localized chemical decomposition left voids. In our experiment, the shock waves may cause similar localized chemical decomposition to create nanoscale voids. Yoffe and Bowden predicted that trapped bubbles of gaseous products could destabilize energetic crystals as their chemistry is activated.[4] A qualitative view of Fig. 4b suggests that the void sizes are ~1 μm in size, but because SEM only probes the material's surface, the image could not be used to quantify the statistics of the void structures. We were unable to zoom into the crystals with higher resolution than that shown in Fig. 4 by SEM, as the focused electron beam decomposed the crystals.

We used USAXS to measure the sub-micrometer void structures in the uncapped sample geometry and to obtain statistically significant, quantitative data about the void structures. USAXS data were acquired using the USAXS instrument at the Advanced Photon Source, Argonne National Laboratory.[51] We measured a representative population of pristine RDX crystals and recovered crystals that had been shocked to maximum pressures of 10 GPa, 16 GPa, and 28 GPa. USAXS measures the scattering intensities produced due to differences (contrast) in electron density. In the case of RDX, after shock, the void structure has very low electron density from any product gases. Hence, shocked crystals produced additional scattering signals that originated from the distribution of void sizes inside each crystal (as compared to USAXS from unshocked RDX). We measured the approximate average size, morphology and volume fraction of the voids by modeling the scattering profiles. We used the information from SEM images for initial fitting parameters (i.e. the mean size, minimum size, volume fraction, and aspect ratio) to model the USAXS data.

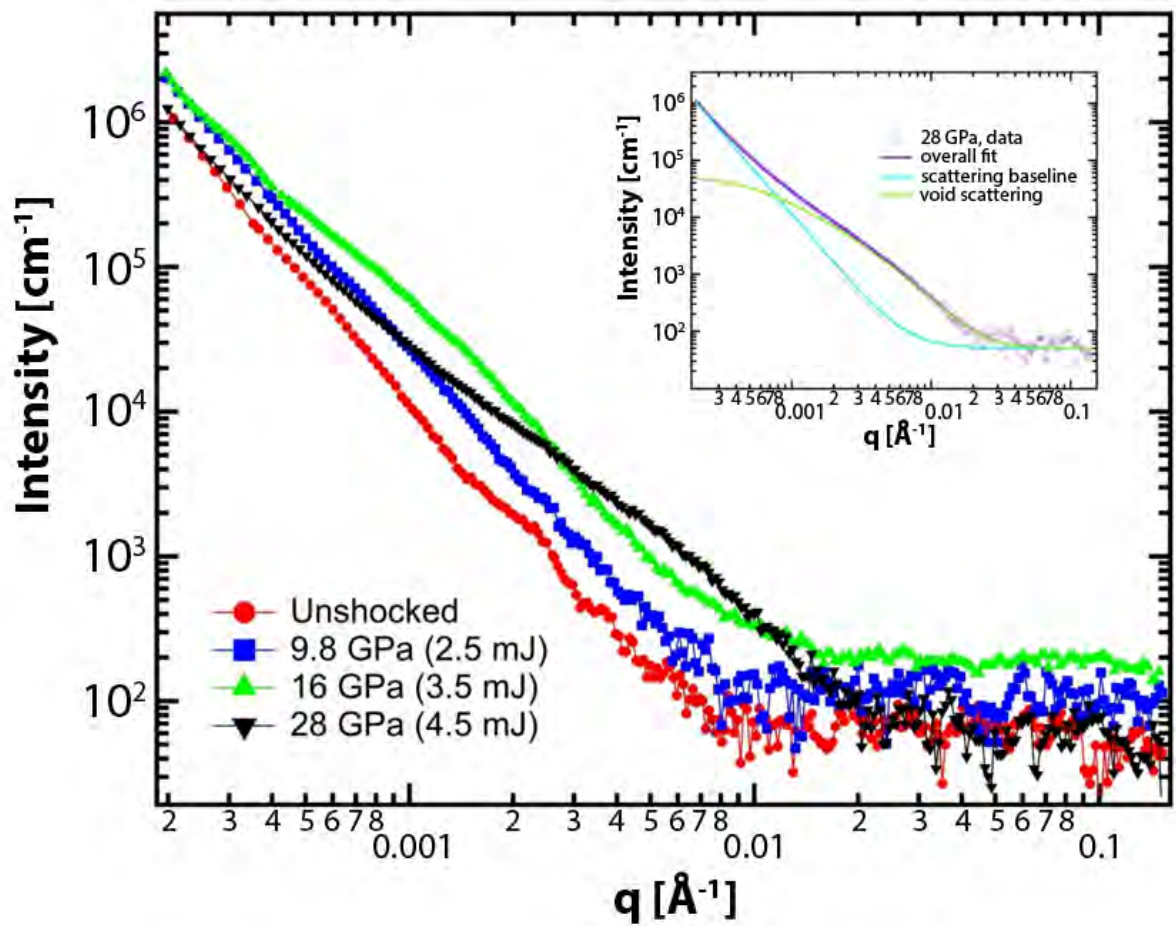

**Figure 5.** USAXS profiles acquired from an unshocked RDX crystal and from representative crystals recovered after being shocked to pressures $P$ = 10 GPa, 16 GPa and 28 GPa in the uncapped sample geometry. All profiles shown here have been corrected for sample thickness and background scattering to provide the absolute scattering intensities.

Fig. 5 shows USAXS profiles of RDX crystals after they were shocked to $P$ = 10 GPa, 16 GPa and 28 GPa as well as an unshocked RDX crystal, all in the uncapped sample geometry. The increase in the scattering intensities of the shocked samples, when compared with that of the unshocked sample, can be attributed to an increase in scattering from void structures like those observed in Fig. 4. USAXS data reduction and analysis for the shocked RDX samples were conducted using the SAS analysis package in Indra and Irena[52], respectively. An example of the model is shown in the inset of Fig. 5. The total scattering intensity is modelled as the sum of a scattering baseline and void scattering. More discussion about this type of scattering model can be found elsewhere.[53] The scattering baseline is constructed based on the scattering signal when the shock-induced voids are not present, i.e., from the unshocked crystal. The void scattering is described by a population of spheroids spatially distributed in the dilute limit. The modeling results are summarized in Table 2.

**Table 2.** Void distributions from shocked RDX samples. In all cases, the scattering profiles were calculated after background subtraction of the baseline scattering contribution, using the calculated scattering cross section of $\sigma$ = 245.5 x $10^{20}$ cm$^{-4}$ and an RDX mass density of 1.799 g/cm$^3$.

| $P$ [GPa] | Mean R$_g$ [nm] | Spheroid Aspect Ratio | Volume Fraction | Number Density |
|---|---|---|---|---|
| 10 | $497 \pm 8$ | 6.1 | $0.0064 \pm 0.0005$ | $(1.8 \pm 0.2) \times 10^9$ |
| 16 | $193 \pm 12$ | 2.1 | $0.0085 \pm 0.0003$ | $(1.4 \pm 0.1) \times 10^{11}$ |
| 28 | $30.8 \pm 12$ | 1.3 | $0.0071 \pm 0.0002$ | $(2.8 \pm 1.0) \times 10^{13}$ |

As shown in Table , the results from the USAXS data indicate changes in the average size, number density and morphology of voids with increasing shock pressure. For increasing shock pressure, the average void size decreases significantly, while the aspect ratio shifts from an oblong geometry at low shock pressures to a nearly spherical shape by $P = 28$ GPa. The pressure-dependent changes in void morphology, not directly measured previously, suggest that increased shock pressure may activate new mechanisms to form voids. The dramatic increase in the number density of voids with increasing shock pressure (100-fold increase for each increment) also suggests a pressure-dependent change in the mechanism that causes the voids to form.

### III.    Time-Dependent Emission from Shocked Samples

To investigate the dynamic origins of the shock pressure-dependent responses in the RDX crystals, we measured the time-resolved shock-induced emission from them. The traces displayed in Fig. 6 show the time-dependent intensity of photoemission produced by shocked samples of RDX (red) as compared to the polymer background (blue) at each shock pressure. We note that any additional emission from adiabatic heating in trapped air and fluorescence from the glass and laser dyes are all included in the polymer background traces. All traces in Fig. 6 show significant photoemission from the polymer layer initially and do not show significant photoemission beyond $t = 50$ ns. Comparing the RDX and polymer emission traces indicates a clear threshold pressure of $P \sim 12$ GPa for RDX emission to occur. While low-pressure shocks produced no RDX emission that was discernable from the background signal, high-pressure shocks generated additional emission from the RDX crystals for $< 50$ ns. Images of the unshocked and recovered crystals showed that all crystals shocked to $P > 8$ GPa showed discernable damage upon recovery. Combining these results demonstrates that damage generated below $P \approx 12$ GPa was linked to short-time photoemission, suggesting a change in the mechanism of damage in that pressure range.

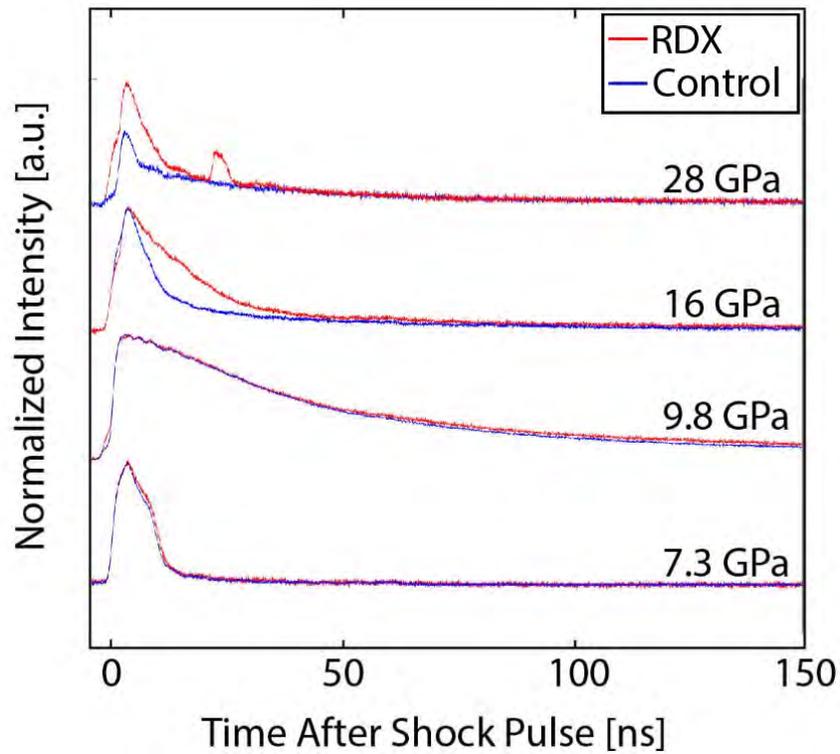

**Figure 6.** Photoemission traces produced by uncapped samples in response to shock waves in our waveguide geometry at four different drive laser pulse energies. Blue traces correspond to representative control emission traces that include the background emission produced by the immobilizing polymer and glass substrates without an RDX crystal, while the red traces include

the RDX crystal. The extra emission peak in the 28 GPa trace is an artifact generated by reflection of stray light from elsewhere in the optical system. The peak emission amplitudes at different pressures are normalized to the same height and do not indicate the relative intensities of signals from one pressure to the next. The amplitudes of emission from samples with and without RDX were normalized to the same levels at long times.

### IV.    *Images of Deformation*

An RDX crystal shocked to a pressure of 10 GPa at the focus is shown in the image sequence in Fig. 7. Parallel deformation planes appear early as striking features and persist throughout the subsequent images; the number of discrete deformation planes and the strength of the depolarization that they produce evolves over the image sequence. The directions, lengths, and propagation of the parallel deformations shown in Fig. 7 are characteristic of the trends we observed for moderate-pressure shocks. As determined by single-crystal X-ray diffraction, this crystal was oriented such that the imaging light propagates normal to the $(\bar{2}10)$ plane. With the 5 ns integration time, the images were unable to resolve the shock wave. The uncollimated imaging light from the diode laser provided high sensitivity to the depolarization induced by the crystal deformations.

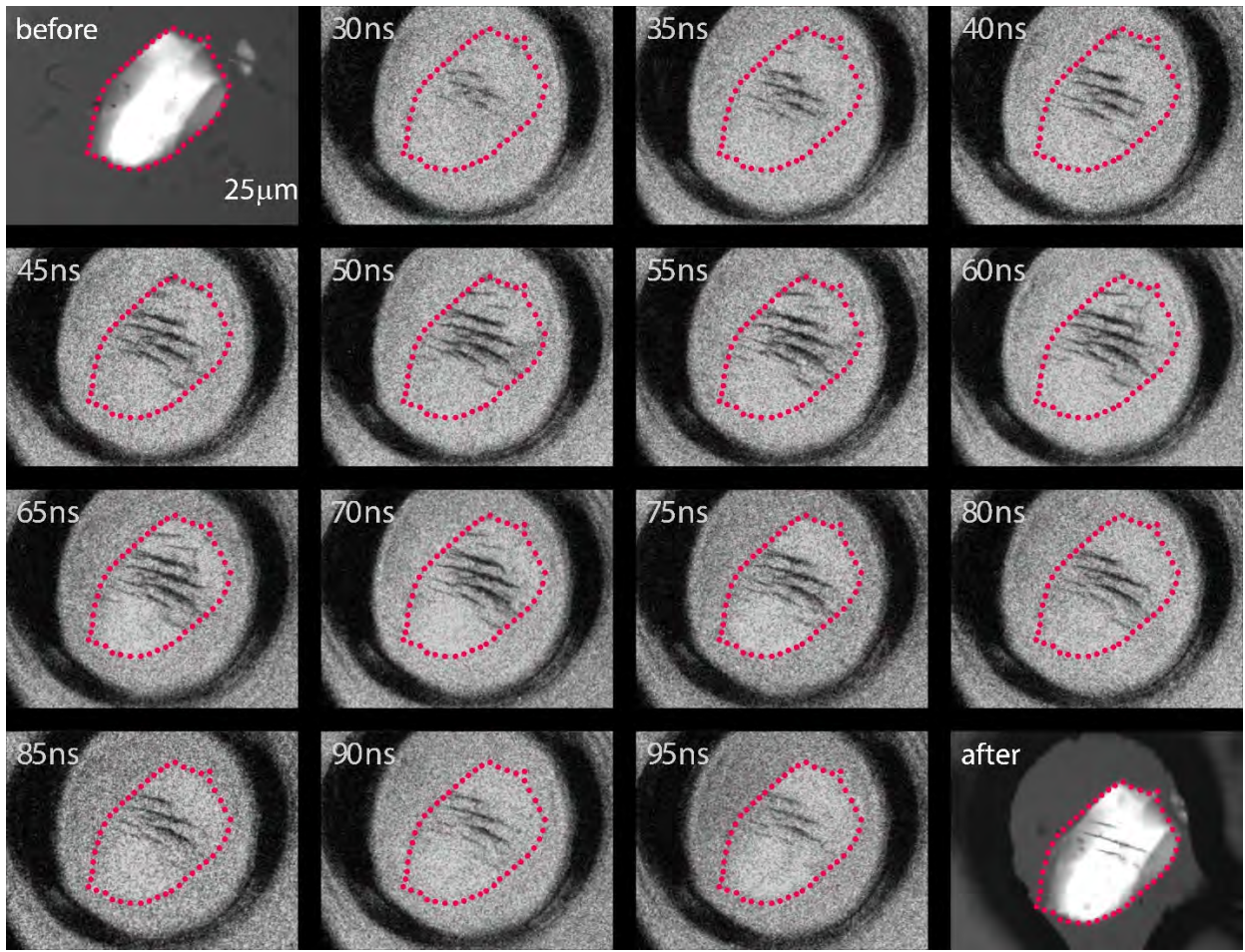

**Figure 7.** Image sequence showing the onset and growth of deformation planes produced by a shock of $P \approx 10$ GPa using $E_{\text{drive}} = 2.8$ mJ. The sequence was collected with the uncollimated quasi-CW imaging light, with a 5 ns integration time for each frame. All images were collected through crossed polarizers. The dotted line indicates the outline of the shocked RDX crystal.

While the shock wave is not visible with the 5 ns integration time, fs-image sequences recorded under similar conditions indicate that the shock reached the center of convergence at $t \approx$ 28 ns. The shock timing indicates that deformations in the crystals began to form within 5 ns of the shock's arrival. For the first 30 ns shown in Fig. 7 ($t < 60$ ns), the initial deformation lines appear, elongate and create new lines all along the same direction. The deformation lines initially grow darker with time, reaching their fullest extent around $t = 65$ ns, and then fade somewhat. Features resembling those in Fig. 7 were never observed in samples with only polystyrene and no RDX present. Neither glass nor polystyrene have long-range order that would introduce reproducibly preferential deformation planes at 5 μm to 50 μm length scales during or after the shock.[54,55] It is clear that the reproducibly parallel features in the images originate from the shocked RDX.

The intensity variation that indicates the growth of the deformation lines is easier to observe with pseudo-color, as shown in Fig. 8a. The linear feature positions and shapes found by locally adaptive discriminant analysis[43,44] correspond to the maps shown in Fig. 8b, providing a clear view of the growth and coalescence of the deformation features.

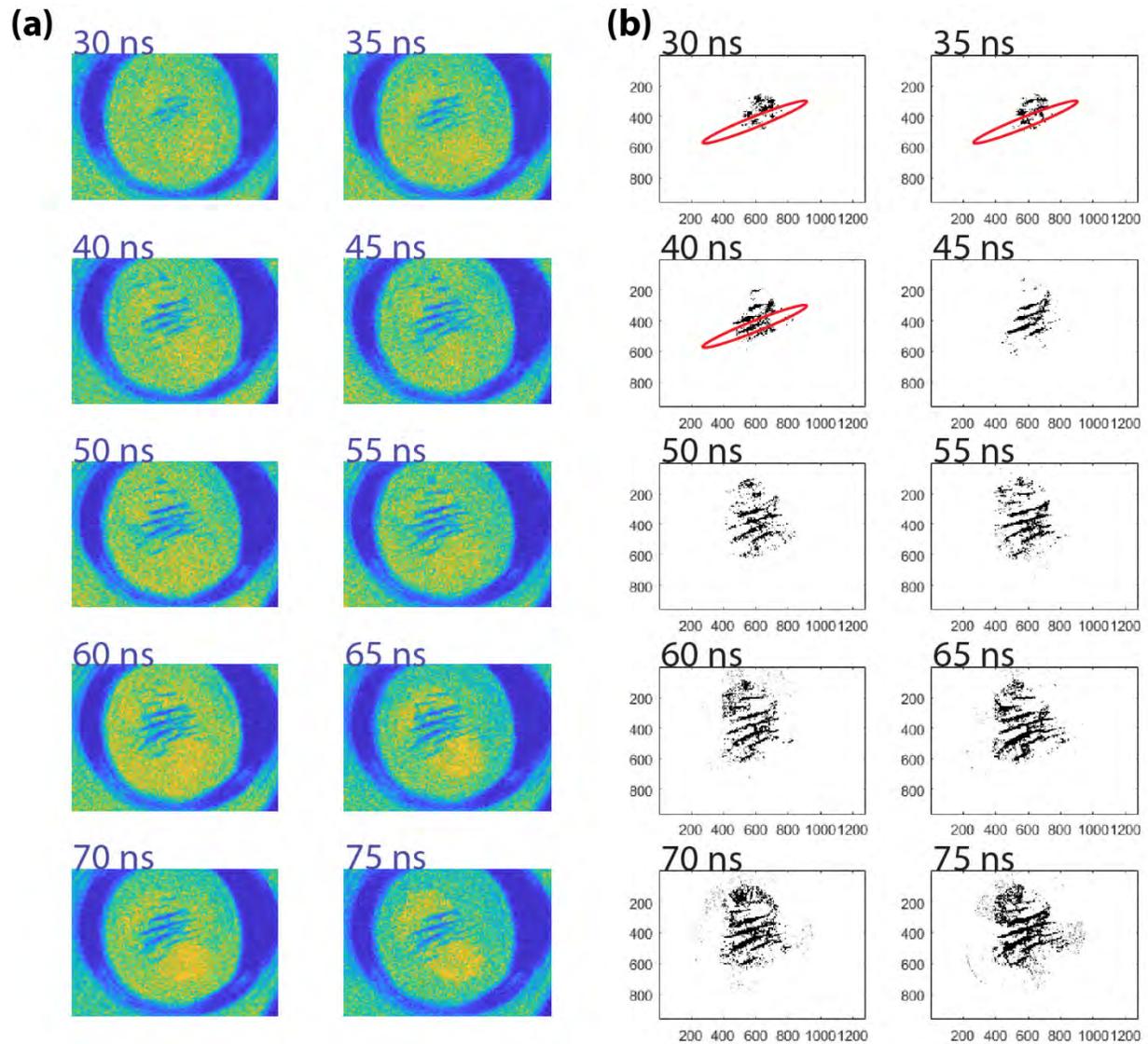

**Figure 8.** (a) A portion of the image sequence displayed in Fig. 7 showing the growth of parallel line features, with pseudo-color to show the intensity variation. These dark features were then located with LADA,[43,44] providing a map of the discrete lines from each frame (b). The red ellipses in the first three frames of (b) show the growth of a single deformation plane.

Fig. 8 demonstrates that the linear deformation planes grow significantly between each 5 ns frame. While each line is quite short at $t = 30$ ns, the lines circled in red grow slightly and coalesce over the first few frames. Similar coalescing features may be seen for other lines across the entire sequence, suggesting that crystallographic planes may have originally been destabilized in multiple places, at smaller length-scales than our ~ 1 μm resolution. At $t = 60$ ns, the bottom pair of deformation lines appear to link along a new direction that is nearly perpendicular to the primary deformation direction.

The deformation map in Fig. 8 demonstrates the characteristic behavior observed in systems with $P < 12$ GPa, which may be seen in other image sequences in the Supplemental Information. Crystals shocked with $P > 12$ GPa, however, produced different deformation dynamics.

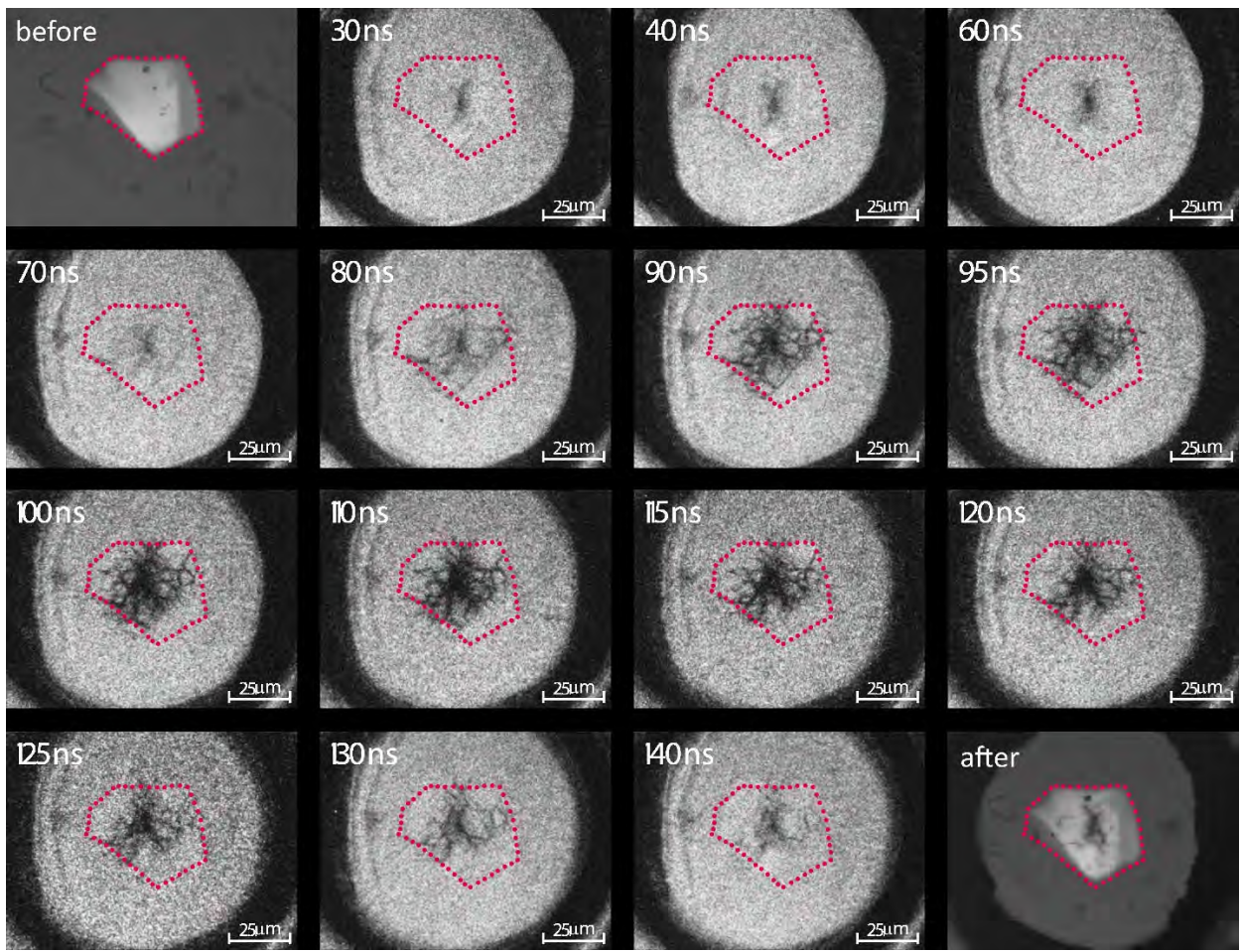

**Figure 9.** Image sequence for an RDX crystal (outlined in red) shocked with a pressure that reaches 28 GPa at the center of convergence (produced by a 4.5 mJ drive laser pulse). The image sequence was collected through crossed polarizers with a 5 ns integration time.

The images in Fig. 9 demonstrate the typical trends for the growth, propagation, and fading of deformation features following the converging shock that reaches considerably higher pressure (~28 GPa) in RDX. While the darkest deformation feature is located at the center of convergence, additional deformation planes span much of the crystal in multiple directions. The image sequence in Fig. 9 illustrates that converging shocks reaching a pressure considerably higher than 12 GPa induce deformations that are much more extensive than in the low-pressure case. We were unable to map the characteristic directions of the high-pressure deformation planes from Fig. 9 because the crossed-polarizers' thresholded high sensitivity produced shadows from all portions of each overlapping deformation lines, blurring our view of each specific line.

To resolve the complicated deformation pathways occurring for shocks of $P > 12$ GPa, femtosecond-resolution image sequences were taken over a 48 ns interval. The 16-frame sequence in Fig. 10 demonstrates the high-pressure deformation progression as resolved by femtosecond shadowgraph imaging. No polarization gating was used. The shadowgraph images clearly resolve the deformation pathways with large spatial variations in $\nabla^2 n$ ($n$ is the refractive index) to resolve the evolution of deformation planes for high pressure shocks.

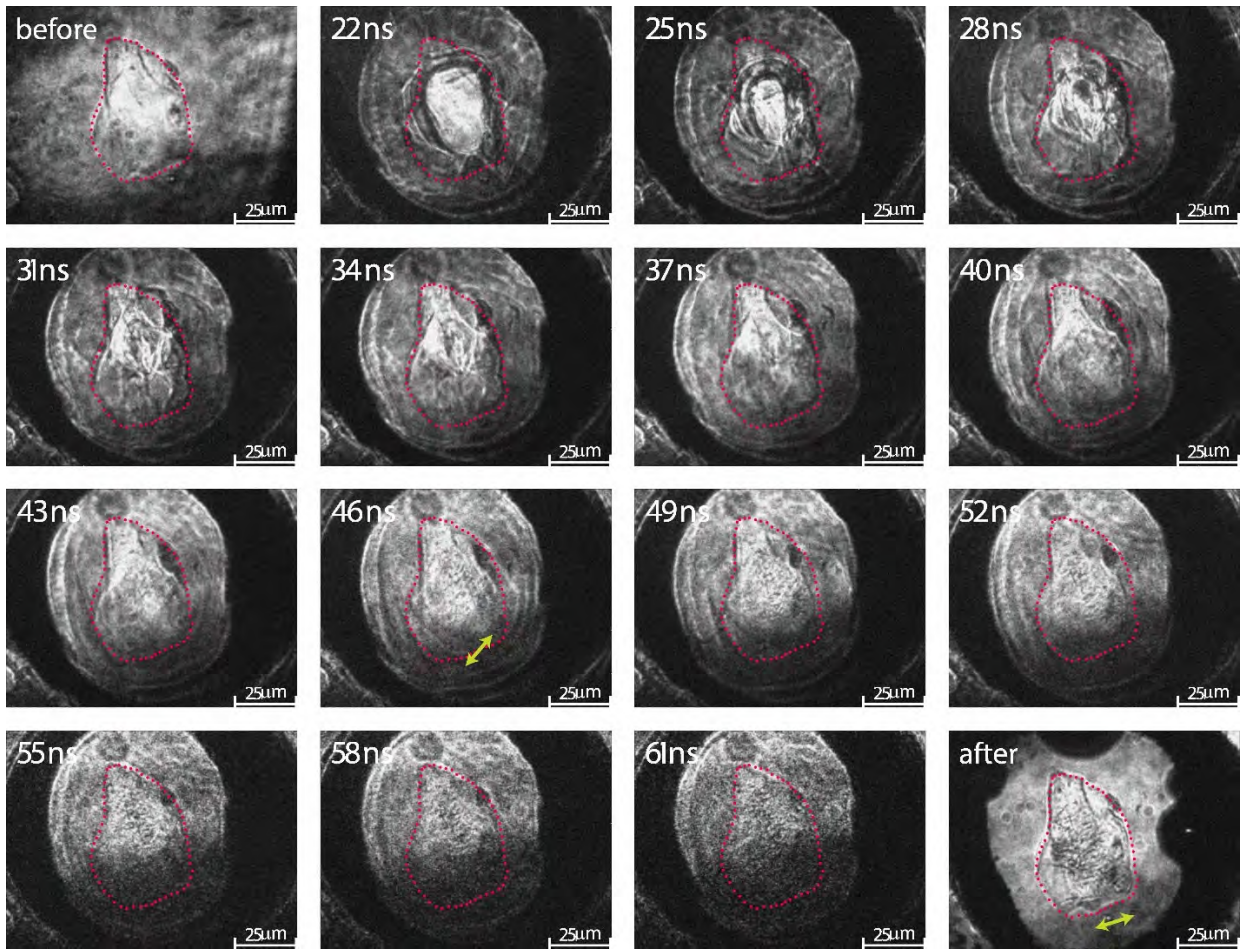

**Figure 10.** The RDX crystal outlined in red is shocked to $P \approx 28$ GPa (with $E_{\text{drive}} = 4.5$ mJ) and monitored using femtosecond multi-frame imaging. The yellow arrows in frames at 46 ns and after show the two deformation directions. Collected without crossed polarizers.

Frames at $t = 22$ ns through $t = 28$ ns in Fig. 10 contain the dark rings corresponding to the converging shock in the polymer. From $t = 28$ ns to $t = 34$ ns, the shock gives rise to complicated image features arising from partial reflections that occur each time the shock reaches the polymer-crystal interfaces[56]. Over this time, the shock converges to a focus and subsequently diverges. By $t = 37$ ns, the shock pressure is too low to produce visible features in the images. All new image features that appear from $t = 37$ ns through the end of the experiment (as well as some of the earlier features) reveal dynamics occurring in the RDX crystal as the material responds to the shock.

Like the images in Fig. 3, the crystal in Fig. 10 reveals indistinct linear features that appear from $t = 40$ ns to $t = 43$ ns and that show clear directional preference parallel to the yellow arrow in the frame at $t = 46$ ns. These indistinct features that appear ~14 ns after the shock entered the RDX crystal are originate from either increased scattering or changes to the refractive index. The linear features in our images originate either from changes to the refractive index or localized scattering of the illuminating light. Changes to the refractive index could indicate either additional compression or phase transitions along these lines, while enhanced scatter at the features would indicate a collection of localized sub-micrometer void structures, similar to those observed in Fig. 3. In this case, given that these features appear and evolve upon release and recovery from the shock pressure, we hypothesize that these features may indicate either nanofractures or accumulating defect pile-ups with <1 μm sizes[57]. As RDX has shown both brittle and ductile responses, the character of these indistinct features is unclear and further experiments are required to identify the cause of the lines. By $t = 46$ ns, the initially poorly-resolved shapes develop into discernable (though diffuse) lines along the same direction that was indicated initially. Similar diffuse linear features following the shock wave were only evident in limited cases (shown in the Supplement), as they depended strongly on the shock pressure and crystal orientation.

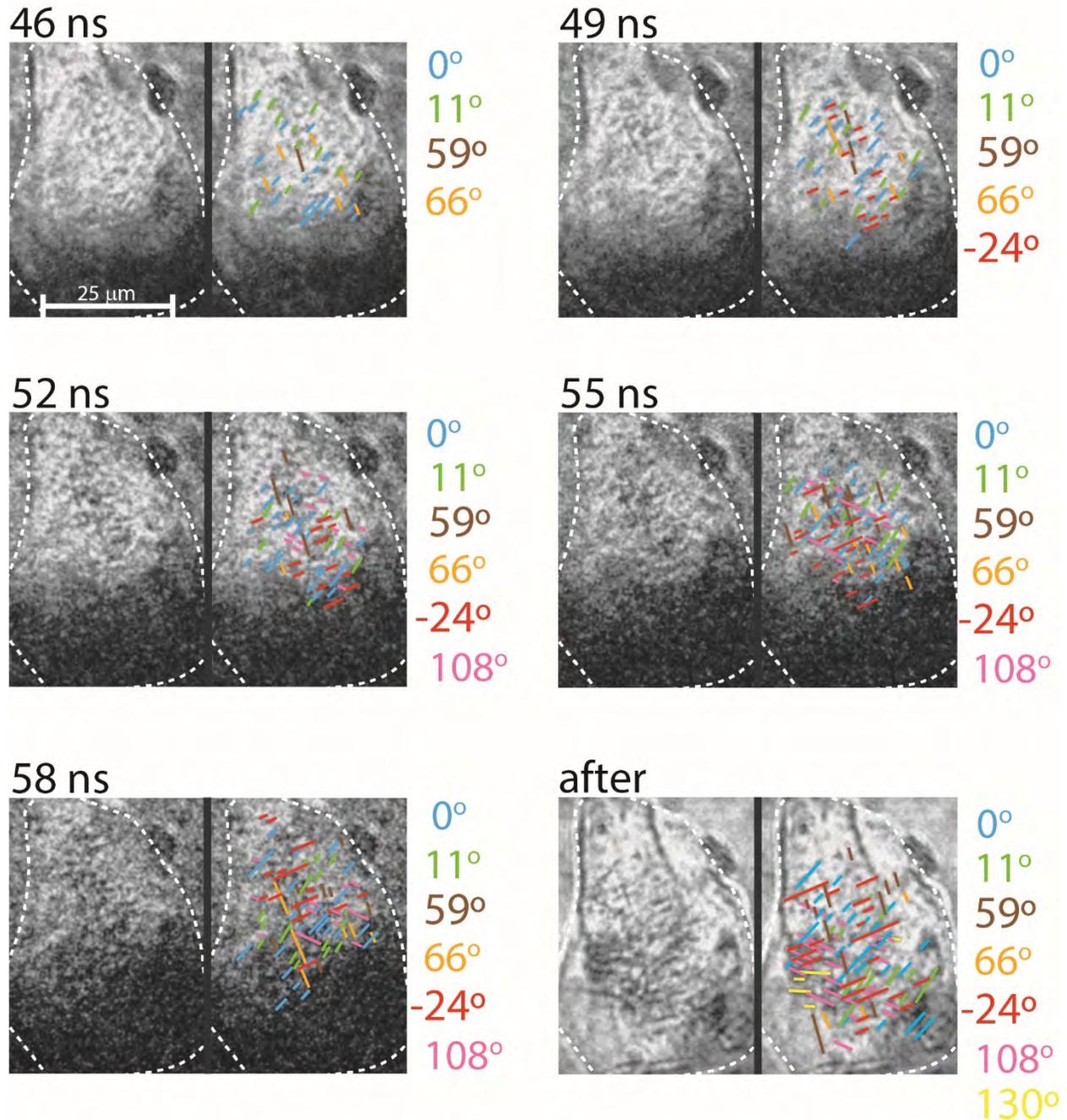

**Figure 11.** Enlarged images from Fig. 10 showing the deformation pathways during a 12-ns interval. Each image is shown annotated (right) and unannotated (left) for each frame. Annotations show the direction of each family of deformation plane with lines of a different color, as labeled on the right. All planes are referenced to the first plane to appear, whose direction we define as 0º.

From $t = 46$ ns until the end of the experiment, discernable lines appear in the crystal that correspond to deformations along specific directions, the primary seven of which are annotated in Fig. 11. We reference the observed angles between each plane to the first plane to appear, highlighted in blue and labelled as 0°. In the first frame in Fig. 11, we observe four families of deformation planes along 0°, 11°, 59° and 66°, with an additional family of planes appearing 3 ns later along -24° and another 3 ns later along 108°. The recovered crystal shows an additional seventh family of deformation planes along 130°, which appeared after the 48 ns window of the experiment.

We do not provide a detailed account of the evolution of the linear features, as their diffuse structure makes quantification rather subjective. However, the annotated and unannotated frames in Fig. 11 demonstrate that both the length and number density of the planes along different directions show different kinetics and different extents of deformation. While the growth kinetics differ between planes, each direction of planes generally appeared as a series of short lines which shifted within the crystal and ultimately coalesced into the final longer deformations.

A close look reveals both evolution and apparent motion of some linear features over the progression of images. Figure 12 displays a further enlarged view of the same RDX crystal, with consistent locations annotated with dashed lines to clarify the changes and motion.

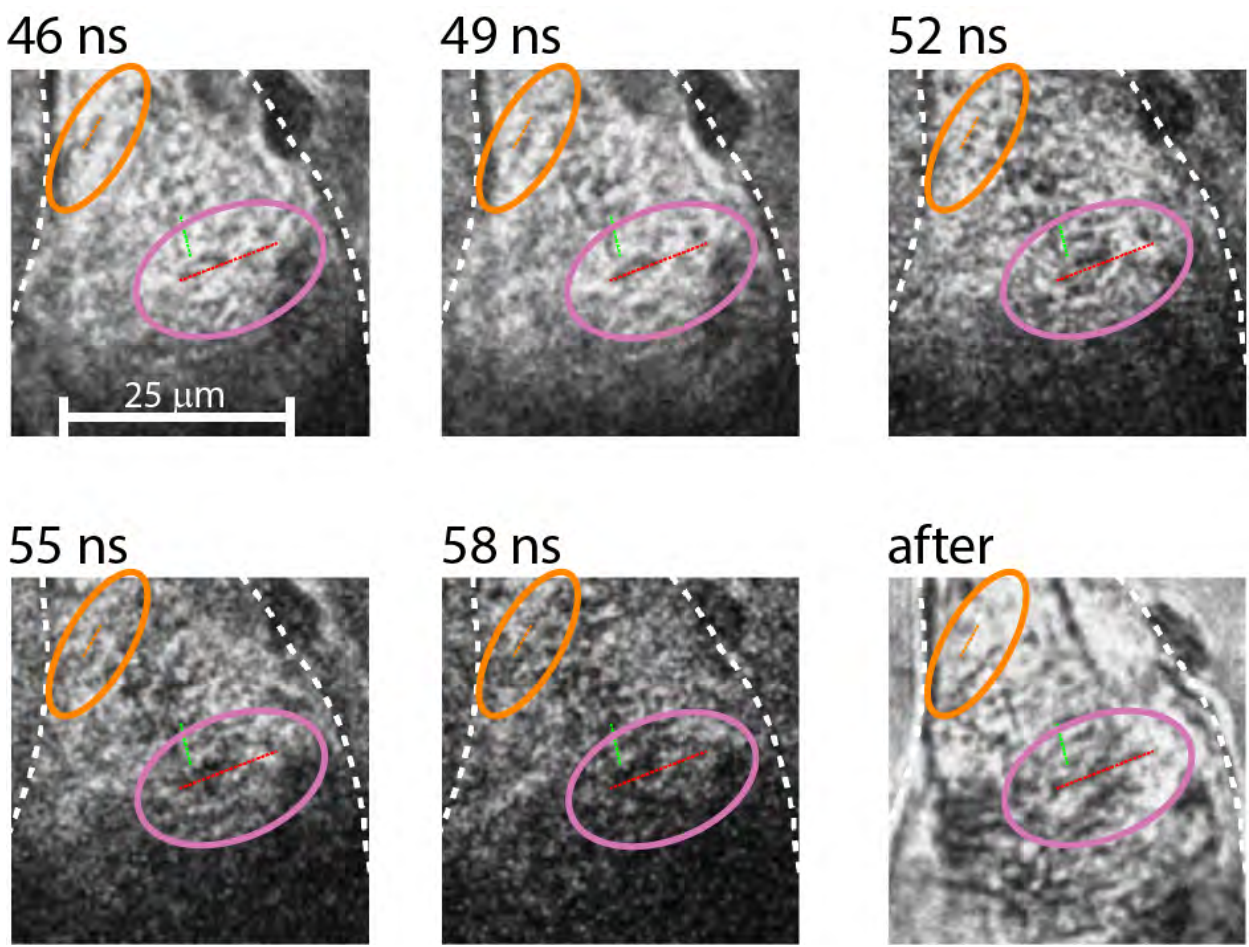

**Figure 12.** Zoomed-in view of deformations in the RDX crystal from Fig. 10 and Fig. 11. The orange, red and green dashed lines indicate locations in which planes evolve along the 0°, 59° and -24° directions, respectively, over the course of the image progression.

The orange dashed line in Fig. 12 (circled in orange) illustrates motion that we observe for some deformation planes, which we detect as a gradual shift in the line over the course of the image sequence. While the dashed line is drawn over the same place in each frame, the deformation plane both moves and lengthens over the image sequence. Altogether the deformation plane migrated ~5 μm by the time the sample was recovered. In contrast, the green dashed line (drawn parallel and to the right of the initial deformation line) demonstrates deformation planes that remain essentially stationary. While this plane moved almost imperceptibly between each subsequent *in-situ* frame, it shifted at longer times $t > 58$ ns.

The red dashed line in Fig. 12 marks the position in each image where a very long feature eventually forms. Frames from 46 ns to 58 ns in Fig. 12 reveal that nearby short lines form and migrate to agglomerate along the red line. As the lines coalesce, the shape of the feature distorts to accommodate the directions of the short linear components, suggesting that differences among the directions of nearby lines do not preclude their interactions. Similar distortion and migration patterns are observable for other planes from the image progression in Fig. 12.

The complicated evolution of the features as they grow—long after the shock has passed— suggests dynamics on some lengthscales that we cannot resolve in our images[58]. Low mobility along some planes suggests different deformation mechanisms along those directions (e.g. high mobility causes plasticity, while low mobility causes fracture[17]). Further detail about the origins of different behavior along different planes requires information about the crystallographic orientation of the crystal, which has not been resolved.

## DISCUSSION

As has been described previously[42,49], the waveguide geometry used herein supports shock waves with a constantly varying pressure. This means that a gradual release of the shock pressure begins immediately following the shock front, without a sustained region of constant high pressure (a.k.a. an unsustained shock). Previous interferometric measurements have determined that the duration of the elevated pressure for these shocks is ~15-20 ns, with most of the duration corresponding to the gradual release of pressure[49]. To understand the RDX shock response in this waveguide system, we use the quantifiable results from the USAXS and photoemission measurements to interpret the deformation trends we observe from our image sequences.

The pressure-thresholded photoemission we observed in the cylindrically shocked RDX demonstrates that a shock with P > 12 GPa is required to have sufficient energy to initiate the radiative response. In RDX, short-lived photoemission may result from either short-lived chemistry[21] or fracture.[59] During decomposition, photoemission is produced by spontaneous emission from electronically excited gas-phase molecules that are produced by the reactions.[60] In contrast, fractoluminescence (termed triboluminescence under vacuum) originates from electrons that are liberated when the crystal cleaves upon reaching its yield stress.[59,61] As the emission we observe from our RDX crystals exceeds that of our control experiments over long timescales above 12 GPa, we hypothesize that the emission originates from chemistry over ~50-ns timescales. Our USAXS data corroborates our hypothesis by quantifying the shapes and volume fractions in recovered crystals. While fractoluminescence would leave behind cleaved crystal surfaces (interfaces) with planar voids (cracks) in our recovered crystals, we only observed oblong voids (nearly planar) for crystals recovered from low-pressure (non-radiative) shocks. In contrast, crystals recovered from high-pressure shock waves produced a higher volume of small spherical voids. We note that our RDX samples were single crystals, i.e. not polycrystalline with multiple domains, but they were far from pristine. Growth defects such as solvent inclusions and others likely contributed to the development of hotspots.

The strong pressure threshold we observe in both USAXS and photoluminescence suggests a change in the mechanism by which the RDX responds to shocks with > 12 GPa. The unresolved shapes in Fig. 3 appeared within ~ 1 ns of the shock front, suggests that the initial damage occurs under compression (i.e. during the 10-15 ns duration of the wave in this experimental

geometry[49]). While further damage likely follows the shock front and pressure release (i.e. under tension[63] or subsequent spall[64], and afterward), the appearance of distinct features within ~1 ns of the shock front demonstrates an initial response preceding rarefaction. From the timescales and void morphologies we observe, we hypothesize that low-pressure shocks ($P < 12$ GPa) initiate non-radiative planar deformations (i.e. fracture or slip), while high-pressure shocks ($P > 12$ GPa) initiate local chemical decomposition.

We hypothesize that the correlated chemistry and very small spherical void structures are indicative of sub-critical hotspots formed by the shock. This model—first proposed by Yoffe[4]—predicts that the temperature and pressure behind the shock front localize around initial defects in an energetic material to initiate decomposition[5,8,10]. The heat produced from the initial reaction may then locally propagate the thermally activated chemistry. As the chemistry in RDX forms predominantly gas-phase products[60], localized chemistry should create small voids with a shape that corresponds to that of the reacted zone. We therefore propose that the emission and void morphology indicate short-lived chemistry at small hotspots in the initial 50 ns following the $P > 12$ GPa shock wave.

For $P < 12$ GPa, we observe immobile deformation planes only along a single direction within the crystal. This suggests that in the absence of resolvable chemistry, low-pressure shocks reveal a single uniquely sensitive direction in the crystal even as the stress is applied by a multi-directional cylindrically converging shock wave. Above the ~12 GPa threshold, we observe different families of deformation planes that appear as the chemoluminescence subsides ($t < 50$ ns). The chemistry occurs near the time that the shock pressure traverses the RDX crystal, creating sub-micrometer sized features in our optical images. Even beyond the timescale of the chemiluminescence, we observe deformations along multiple directions—after the crystal has been successively compressed and released twice and heated to an elevated shock temperature. As previous work has demonstrated that hotspots may be formed by dislocation pileup[11], detailed X-ray characterization is required to determine what types of defects and specific crystallographic axes are inherent to the "deformation planes" we observe optically. Given the relevant timescales for the optical deformation features, we hypothesize that the deformations may either be caused by a pileup of nanoscale deformations or by localized pressure from pockets of product gases trapped in the crystal.

We note that phase transitions to the γ or ε phases have been observed for shocks with P > 4 GPa. All of our shock pressures in this work have exceeded 4 GPa, and our observed pressure threshold for chemical decomposition does not match the known phase transition pressure. Considerations such as the kinetics of phase transitions and reversion to the original phase after pressure release, over-driving of phase transitions at high pressures[61,62], and the variation of phase transition behavior as a function of shock propagation direction warrant further study and are beyond the scope of this work

While additional detail is required to fully understand the detailed mechanisms of the changes we observed, our results show reproducible dynamics as RDX crystals embedded in a polymer binder respond to a non-planar shock wave. As shock reflections in application-relevant PBX composites are known to change the shock geometry[12], detailed experiments are required to investigate how changes to the shock geometry influence the resulting chemistry. Uniaxial shocks along crystalline symmetry directions may be described by single strain tensor component, however converging waveguide shock waves include additional stress components that exert different stresses upon the RDX crystals. Inhomogeneity in the shock direction for converging shocks create additional instability, making a direct comparison between uniaxial and waveguide converging shocks difficult.

Previous investigations have shown that uniaxial shocks induce fracture or deformation in RDX and HMX, with crystallographic planes that are activated depending on their orientation with respect to the shock direction[16,25,26,28,67]. This occurs because each slip system requires sufficient stresses to activate and cause fracture, slip or shear banding, which requires the shock stress to have sufficient components along specific crystallographic directions. Our experimental geometry is more representative of the stresses exerted in real PBX formulations, in which small crystallites are randomly oriented and in contact with a polymer binder, leading to complex and turbulent shock geometries. The deformations we observed in RDX suggest that nonplanar shocks cause the crystals to deform first along well-defined sensitive plane before deforming along additional directions. Despite the complexity of the initial stresses, direct real-time imaging of how individual crystals respond to the shock collected with real-time emission and characterization of the void distributions in the recovered crystals has revealed a dominant preferential response in RDX crystals at moderate pressures.

CONCLUSIONS

This work demonstrated that converging shock waves in RDX produce plasticity and chemical activation that are strongly thresholded by the shock pressure. For low-pressure shocks with $P < 12$ GPa, we observe nonradiative shock responses that form roughly planar nanovoids that grow along a single direction. High-pressure shocks that reach pressures above 12 GPa result in photoemission from RDX decomposition products followed by the onset of plasticity or fracture along several families of deformation planes, with small spherical voids that are observed after recovery.

At all pressures, the directional specificity of deformation planes does not correlate to the directions of applied stress in the crystal, suggesting inherent lattice sensitivities in RDX. Above the pressure threshold to activate chemistry, we observe a cascade of deformations along up to seven different directions, with evolution and partial coalescence over tens of nanoseconds that further evolved after our real-time measurements before reaching the final recovered state. The timescale of the high-pressure deformation dynamics and photoemission suggest that the deformations may be driven by evolving forces within the crystal including stresses arising from formation of trapped gaseous reaction products. Our observations of the dynamic responses in RDX to cylindrical shock waves that exert stresses along many crystallographic directions provide new insight into the responses of randomly oriented crystals to geometrically distorted shock waves that occur in real-world PBX formulations.

ASSOCIATED CONTENT

Supplemental Information includes a description of further experimental details and additional image sequences to show statistics of the phenomena described here.

AUTHOR INFORMATION

Corresponding Author: Leora E. Dresselhaus-Cooper, dresselhausc1@llnl.gov

† Present address: [c] Physics Division, Lawrence Livermore National Laboratory, 7000 East Ave.

L-487, Livermore, CA 94550

AUTHOR CONTRIBUTIONS

The project was directed by KAN. Shock experiments were carried out by LEDC. Samples were prepared by LEDC and DJM. USAXS experiments were performed with LEDC, FZ, JI and DJM. The orientation of the crystal in Fig. 7 was measured by LEDC, CT, SYGW and YSC. Analysis of the results were carried out by LEDC and KAN. The manuscript was written by LEDC, with supporting contributions from all authors.

ACKNOWLEDGMENTS

The authors acknowledge William DiNatale, Peter Mueller, Charlie Settens, and Steven Kooi for their instruction and guidance in using MIT instrumentation facilities. The authors also acknowledge Jon Eggert, Betsy Rice, Brian Barnes and Lara Leininger for their conversations that helped guide some of the analysis for this work. The work was supported by the Office of Naval Research grants N00014-16-1-2090 and (DURIP) N00014-15-1-2879. This research used resources of the Advanced Photon Source, a U.S. Department of Energy (DOE) Office of Science User Facility operated for the DOE Office of Science by Argonne National Laboratory under Contract No. DE-AC02-06CH11357. NSF's ChemMatCARS Sector 15 is supported by the National Science Foundation under grant number NSF/CHE-1834750. LEDC while at LLNL were supported under the auspices of the U.S. Department of Energy by Lawrence Livermore National

Laboratory under Contract DE-AC52-07NA27344. LEDC's work was supported in part by LLNL, Lawrence Fellowship Program.

## SUPPORTING INFORMATION

The Supporting Information is available free of charge on the ACS Publications website at DOI:

- Additional experimental details for imaging and scattering experiments are included to detail how the analysis corroborates our interpretation. Additional image sequences are also included to display the trends in our data.

---

[i] Certain commercial instruments, materials, or processes are identified in this paper to adequately specify the experimental procedure. Such identification does not imply recommendation or endorsement by the National Institute of Standards and Technology, nor does it imply that the instruments, materials, or processes identified are necessarily the best available for the purpose.

# Supplemental Information for Publication

# Pressure-Thresholded Response in Cylindrically Shocked Cyclotrimethylene Trinitramine (RDX)

Leora E. Dresselhaus-Cooper,*[ab†] Dmitro Martynowych,[ab] Fan Zhang,[d] Charlene Tsay,[e] Jan Ilavsky,[f] SuYin Grass Wang,[g] Yu-Sheng Chen,[g] Keith A. Nelson[ab]

[a] Department of Chemistry, Massachusetts Institute of Technology, 77 Massachusetts Ave., Cambridge, MA 02139; [b] Institute for Soldier Nanotechnology, Massachusetts Institute of Technology, 77 Massachusetts Ave., Cambridge, MA 02139; [†] Present address: [c] Physics Division, Lawrence Livermore National Laboratory, 7000 East Ave. L-487, Livermore, CA 94550; [d] Materials Measurement Science Division, National Institute of Standards and Technology, 100 Bureau Dr., Gaithersburg, MD 20899; [e] Department of Chemistry, University of California Riverside, 501 Big Springs Rd., Riverside, CA 92521; [f] X-ray Science Division, Argonne National Laboratory, 9700 S. Cass Ave., Argonne, IL 60439; [g] ChemMatCARS, Center for Advanced Radiation Sources, The University of Chicago, 9700 S. Cass Ave., Argonne, IL 60439

> *S1 – Analysis of Crystals Recovered from Shocked Sample*
>
> *S2 – Ultra-Small Angle X-ray Scattering (USAXS)*
>
> *S3 – Imaging Details*
>
> *S4 – Partially Confined Shocks in the Uncapped Sample Geometry*
>
> *S5 – Timing Calibration for Converging Shocks*
>
> *S6 – Supplemental Image Sequences for Capped Sample*
>
> *S7 – Supplemental Image Sequences for Thresholded Dynamics*

*\* Corresponding Author, dresselhausc1@llnl.gov*

### *S1.* *Analysis of Crystals Recovered from Shocked Sample*

The shocked RDX crystals remained available to use for various types of recovery analysis. We conducted scanning electron microscopy, ultra-small angle X-ray scattering (USAXS) and X-ray diffraction to gain information about shock-induced structural damage and crystallographic orientation.

USAXS allowed us to probe statistical populations of void structures ranging from 1 μm to 1 nm in size in shocked RDX crystals using 24 keV X-rays ($\lambda = 0.517 Å$) at Beamline 9-ID-C at the Advanced Photon Source.[1] The USAXS instrument, which makes use of multiple reflections off perfect crystal optics to access the angular space at and near the forward scattering direction of the x-ray beam, provides a primary calibration of the absolute scattering intensity.[2] The X-rays propagated normal to the sample plane, along the same path as the shock drive and imaging beams shown in Figure 1, illuminating a square 200 μm x 200 μm spot that completely covered each crystal of interest. Background subtraction of USAXS profiles measured from the blank glass substrates and measurement of the sample thicknesses by confocal microscopy permitted determination of the absolute intensity per unit volume of RDX. For each shock pressure, USAXS profiles from at least four different shocked RDX crystals were collected to ensure that the data represented the characteristic structure of crystals recovered from the given conditions.

All initial guesses for the USAXS profiles were guided by scanning electron microscopy (SEM) images collected from the capped samples. The non-conducting organic samples were coated with 10 nm of gold, and a Jeol 6010LA microscope, with an accelerating current of 5 kV to 15 kV, was used to construct electron microscopy images with 800x to 1400x magnification. Higher resolution and magnification were not attainable in the instrument without damaging the samples.

X-ray diffraction was performed to determine the orientation of one of the recovered crystals at 15-ID-D in the Advanced Photon Source. For the immobilized crystals inside the fused silica plates, penetrating the fused silica windows required hard X-ray photon energies that were higher than the core electronic resonance of the Si Kα peak (centered at 1.5 keV).[3] A > 150 μm diameter beam of 30 keV monochromatic X-rays irradiated the entire crystal, with the incident beam propagating along the same axis as the shock drive and imaging probe beams shown in Figure 1 for $\omega = 0°$, $\phi = 0°$ and $\chi = 0°$. We collected a sequence of diffraction patterns by scanning a goniometer across a range of -15° to 15° in each angular dimension (in 0.5° increments) and we used the Bruker Apex software to solve for the orientation matrix. With this information, we then used the software on the Bruker apparatus at the MIT X-ray diffraction lab to determine the crystallographic orientation of the incident face of the crystal (i.e. to face index) the crystal.

### S2.    *Ultra-Small Angle X-ray Scattering (USAXS)*

The small-angle X-ray scattering (SAXS) experiments probe the morphology of mesoscopic structural heterogeneities and determine their sizes, shapes, and volumes.[4] USAXS, as a special case of SAXS, examines structural heterogonies at a larger length scale (up to 20 μm). When the scattering intensity originates from a particulate system, so long as the scattering intensity is absolutely calibrated, a proper SAXS analysis can give the particle size distributions and volume fractions for particles with known contrast.

SAXS analysis relies on using existing knowledge or evidence of the materials system to construct scattering models to fit the observed scattering profile.

For small angle scattering experiments conducted at a synchrotron, the polarization factor is usually negligible. As the scattering angle $2\theta$ depends on the X-ray wavelength, scattering is often described by the wavelength-independent scattering wavevector $q = 4\pi \sin(\theta)/\lambda$. The scattering contrast creates additional scattering contributions from the individual scatterers, described by

$$\Delta I_1(q) = I_0 \, \Delta\rho^2 \, V_1^2 \, P(q) \qquad (1)$$

where $\Delta I_1(q)$ is the additional scattered intensity from a particle (subtracted from the background scattering) with volume $V_1$ and electron density difference $\Delta\rho$ from that of the matrix. The form factor $P(q)$ describes the size, shape and electron density of the scattering particle,[4] often described by $P(q) = [\Delta\rho \int \cos(\boldsymbol{q} \cdot \boldsymbol{r}) \, dV]^2$ where the integral is over the particle volume. For an ensemble of $N$ scattering particles, we see that the total additional scattering intensity from all the particles is given by

$$\Delta I(q) = N\,\Delta I_1(q)\,S(q), \tag{2}$$

where $S(q)$ is the structure factor, which describes the scattering positions relative to each other. For scattering systems in the dilute limit, i.e., the scattering objects sufficiently far away from each other, $S(q)$ is often approximated as 1.

Analysis of the form and structure factors gives specific information about the shapes and sizes of the scattering particles, as well as their separations. This forms the basis of small angle scattering. In practice, small angle scattering analysis can be difficult, mostly due to its nature of being an inverse problem. Construction of a good scattering model relies on pre-existing knowledge about the scattering system. The present case is favorable in that the "particles" of interest are voids filled with air, with negligible electron density, so only the RDX scattering intensity needs to be determined. We measured the background scattering intensity of an unshocked pristine crystal of RDX, allowing us to obtain the relevant functional form of $I_0$ for the RDX "matrix." With this scattering baseline, we then fit the additional scattering intensity $\Delta I(q)$ using a spheroid model in the Irena package of Igor (Wavemetrics), developed for SAXS analysis.[5] This combined model allows us to quantify the mean and variance of the number density, sizes and shapes of the voids within the closed samples.

### S3.    *Imaging Details*

Image sequences were collected with two different imaging configurations to measure dynamics over different timescales. To fully resolve the shock wave, we used a multi-frame single-shot imaging configuration described elsewhere.[6] The 130 fs pulse duration of each imaging frame set the integration time for each frame. Over that duration, even a fast shock of $U_S = 20$ km/s could only traverse 2.6 nm during each image, which is far smaller than our ~ 1 μm optical resolution. With 130 fs temporal resolution, this technique was able to make the shock appear stationary in each frame.

The deformation dynamics that followed the shock occurred over a significantly longer duration than was accessible in 16 frames using the fs-resolution imaging technique. To image the slower crystal deformations, we used a 10 μs SILUX diode laser system (λ = 610 nm) to illuminate the sample. The electronic gating in the multi-frame camera was used to set variable time intervals and integration times ($\geq$ 5 ns duration) for the long-timescale image sequences.

Both imaging configurations used the intensified CCDs in the Specialized Imaging 16X Camera to measure each image sequence, with temporal jitter of $\pm 1$ ns on each shot. Additional timing drift occurred over the course of minutes or hours but was accounted for by manually adjusting the global delay during the experiment and ensuring that the first frame was always dark to validate that the first image came from the first pulse in the pulse train. All image sequences used shadowgraph imaging, which is sensitive to the Laplacian of the refractive index.[7] The images produced bright and dark features corresponding to regions with sharply varying position-dependent refractive index.

### S4. *Partially Confined Shocks in the Uncapped Sample Geometry*

In Figure 2 from the manuscript, the feature that appears to show a shock wave in the RDX crystal originates from a shock component in the glycerol capping fluid[6]. The same clear wave cannot be unambiguously defined without the presence of the capping fluid, as is shown in Figure S1. In this figure, the cylindrical rings corresponding to the shock distort slightly, but broaden significantly as the shock traverses through and over the crystal. As the RDX crystal is thinner than the full sample layer, the images in Figure S1 should include shock features in the RDX crystal and both polymer layers surrounding it.

As the shock features in shadowgraph images originate from $\nabla^2 n$, features can only be clearly resolved when fast and large spatial changes in the refractive index occur within the material. This causes materials with high photoelastic constant to show strong features, while those with low photoelastic constants can be difficult to resolve[6]. While the photoelastic response of RDX has never been measured, its lack of discernable shadowgraph features during a shock wave suggests that the photoelastic response is quite small.

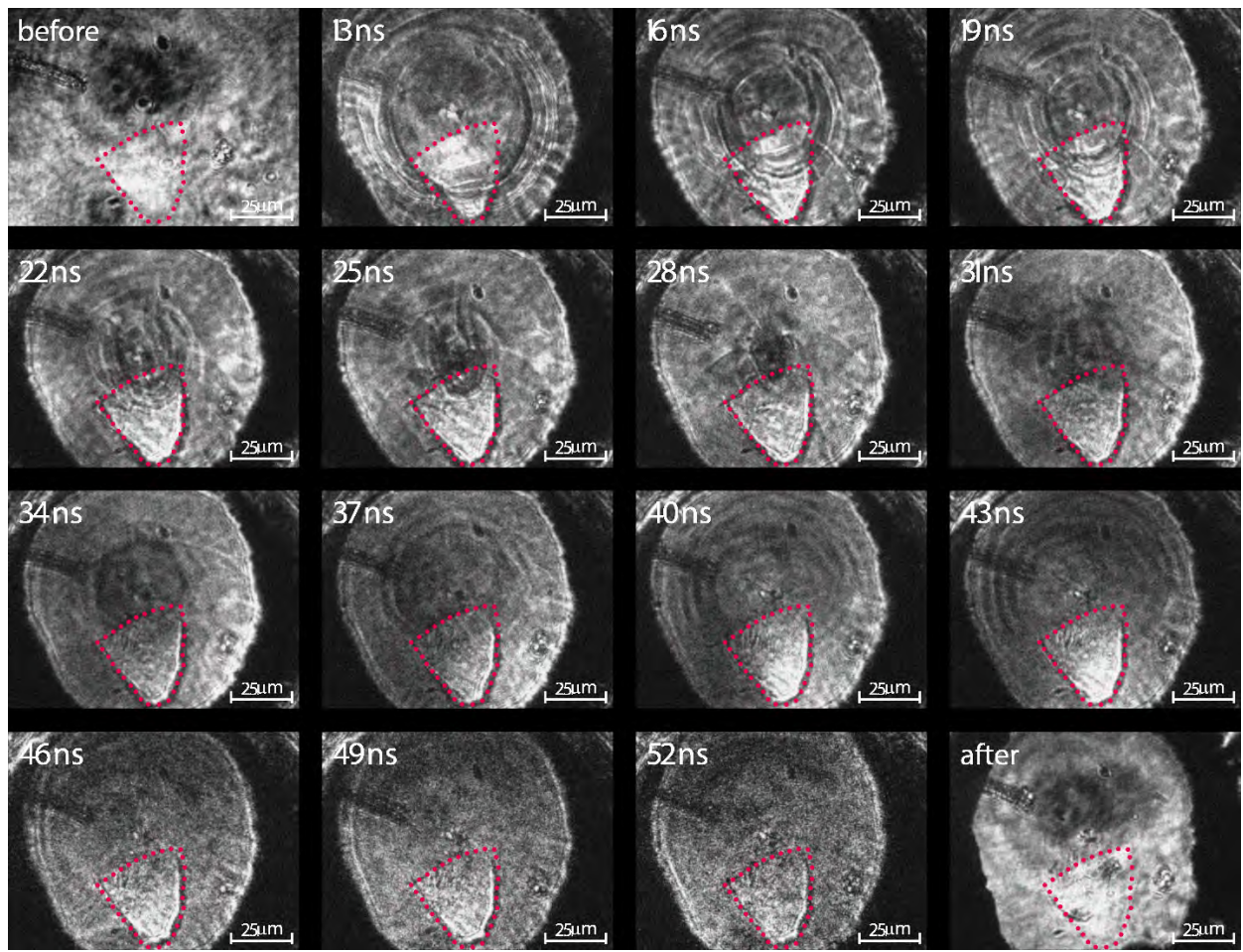

**Figure S1.** *Image sequence showing the shock wave traversing the RDX crystal upon convergence and divergence for a P = 16 GPa shock (E$_{drive}$ = 3.3 mJ). Images from this sequence were collected using shadowgraph imaging, by the fs probe technique described in the main paper.*

Measuring the shock pressure inside the RDX crystals requires either the velocity of the wave or the compressed material density. While the uncapped samples had clear and unambiguous signals for the shock position (coming from the capping fluid), the complicated *P-T* states of the uncapped samples make their shock velocities poor indicators of the states reached in the capped samples.

### S5.    *Timing Calibration for Converging Shocks*

In Fig. S2, the shock converges and traverses the crystal from $t < 13$ ns until $t = 25$ ns to 28 ns, after which it diverges. The shock appears near the center of convergence in the frames collected at 25 ns and 28 ns, however, the precise timing at which the shock focuses is not resolvable. As has been shown previously,[6] the converging shock waves generated in this waveguide geometry produce shocks that have a coupled wave structure between the substrates and the samples. The additional shock components in the coupled wave structure ultimately cause the converging shock to reach its focal point at a range of different times. This and other similar image sequences were used to specify our timing expectations for shocks at $P = 16$ GPa.

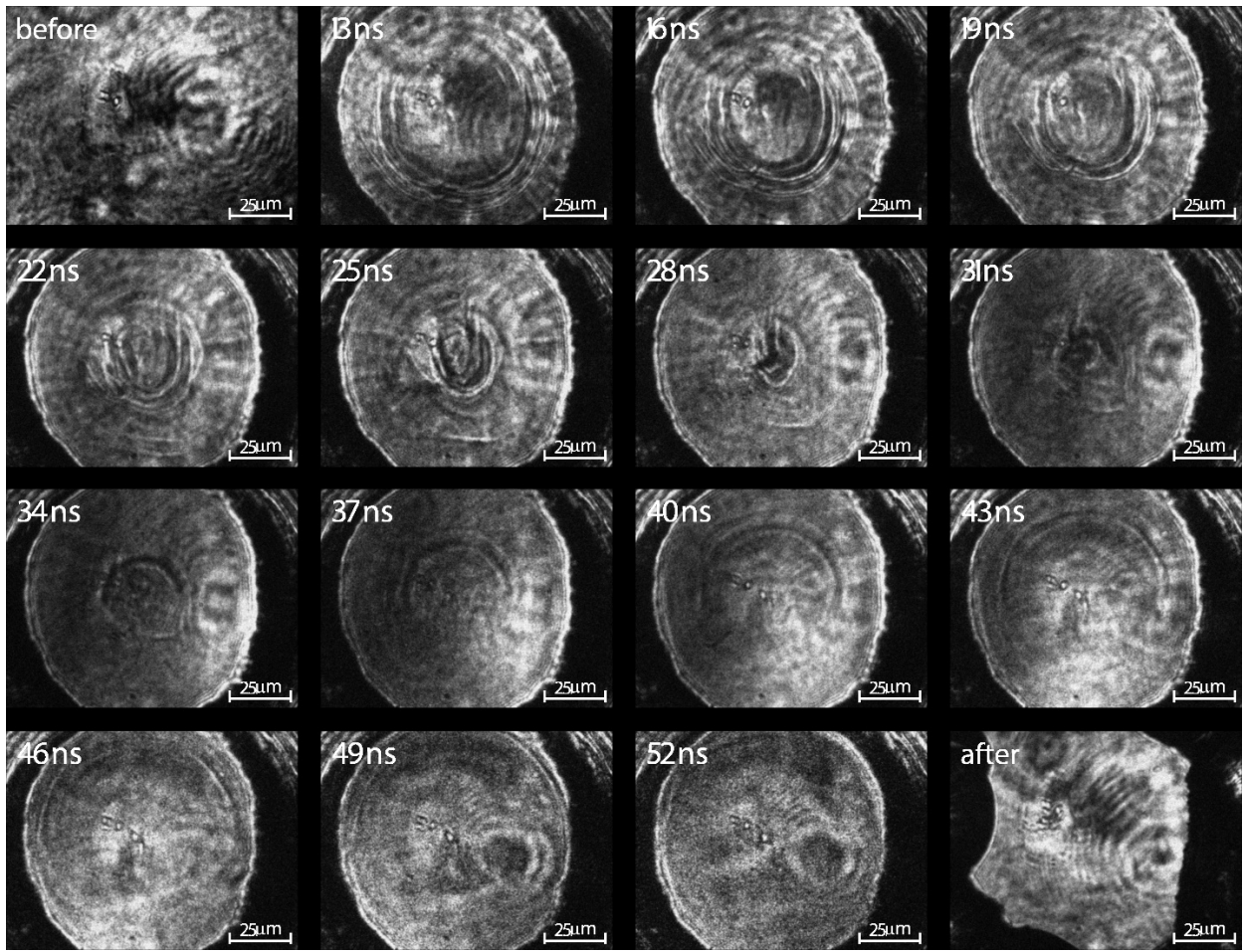

**Figure S2.** *Image sequence showing the shock wave traversing the RDX crystal upon convergence and divergence for a P = 10 GPa shock ($E_{drive}$ = 2.4 mJ). Images from this sequence were collected using shadowgraph imaging, by the fs probe technique described in the main paper.*

Fig. S2 presents a similar image sequence for the $P \sim 9$ GPa shock. In this image sequence, the shock arrives at the center of convergence in the frame at $t = 28$ ns. The details from this timing are used in the main manuscript to assess how long it takes for the shock to induce deformations at the micron length scale.

### *S6.*     *Supplemental Image Sequences for Capped Sample*

To demonstrate that the shocked crystals from Fig. 2 in the main paper show a converging shock wave, we include a supplementary image from a similar experiment (Fig. S3). This image sequence shows a shock produced by the same laser energy (corresponding to $P \sim 16$ GPa), clearly revealing convergence. The image sequence in the main text does not include as many frames during convergence but one frame in particular shows the shock clearly overlapped with the RDX crystal. In our experiments the image contrast and the timing of shock arrival at an RDX crystal varied from shot to shot, but a multitude of measurements with and without RDX crystals present showed that comparable excitation conditions resulted reliably in comparable shock responses. We also note that the capped samples were not used for most of our experiments, as the additional interfaces caused imaging artifacts that were difficult to interpret.

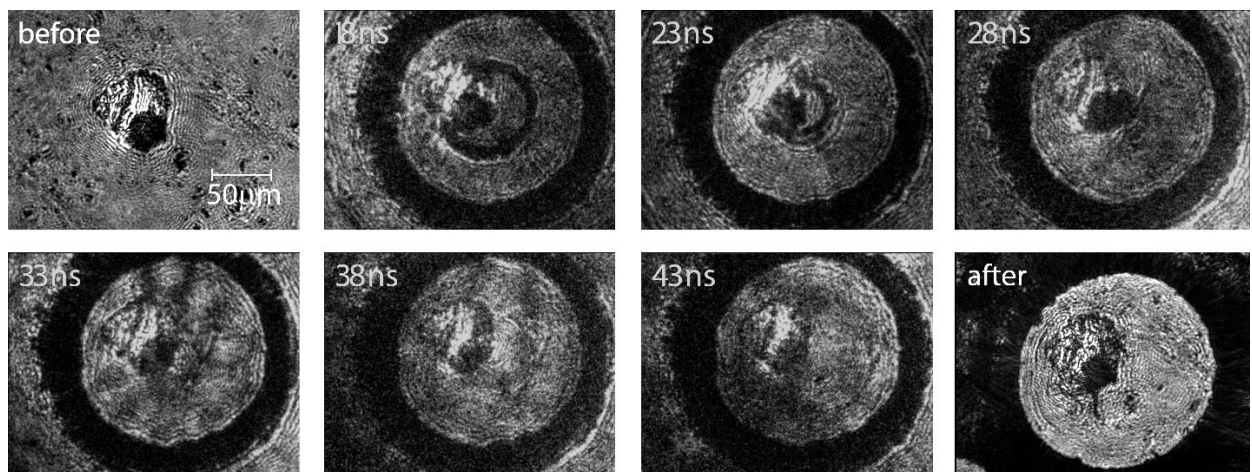

**Figure S3.** *Image sequence collected with the capped sample geometry, using 3.5 mJ of drive laser energy (P ~ 16 GPa), showing the converging shock wave under these sample conditions.*

### *S7.*     *Supplemental Image Sequences for Thresholded Dynamics*

The thresholded deformation dynamics described in this paper were generally reproducible. While the specific details of the deformation features, timescales and directions differed for each crystal, the effects were largely consistent for measurements conducted with the same drive pulse energy. The image sequences given below demonstrate the reproducibility of the effects.

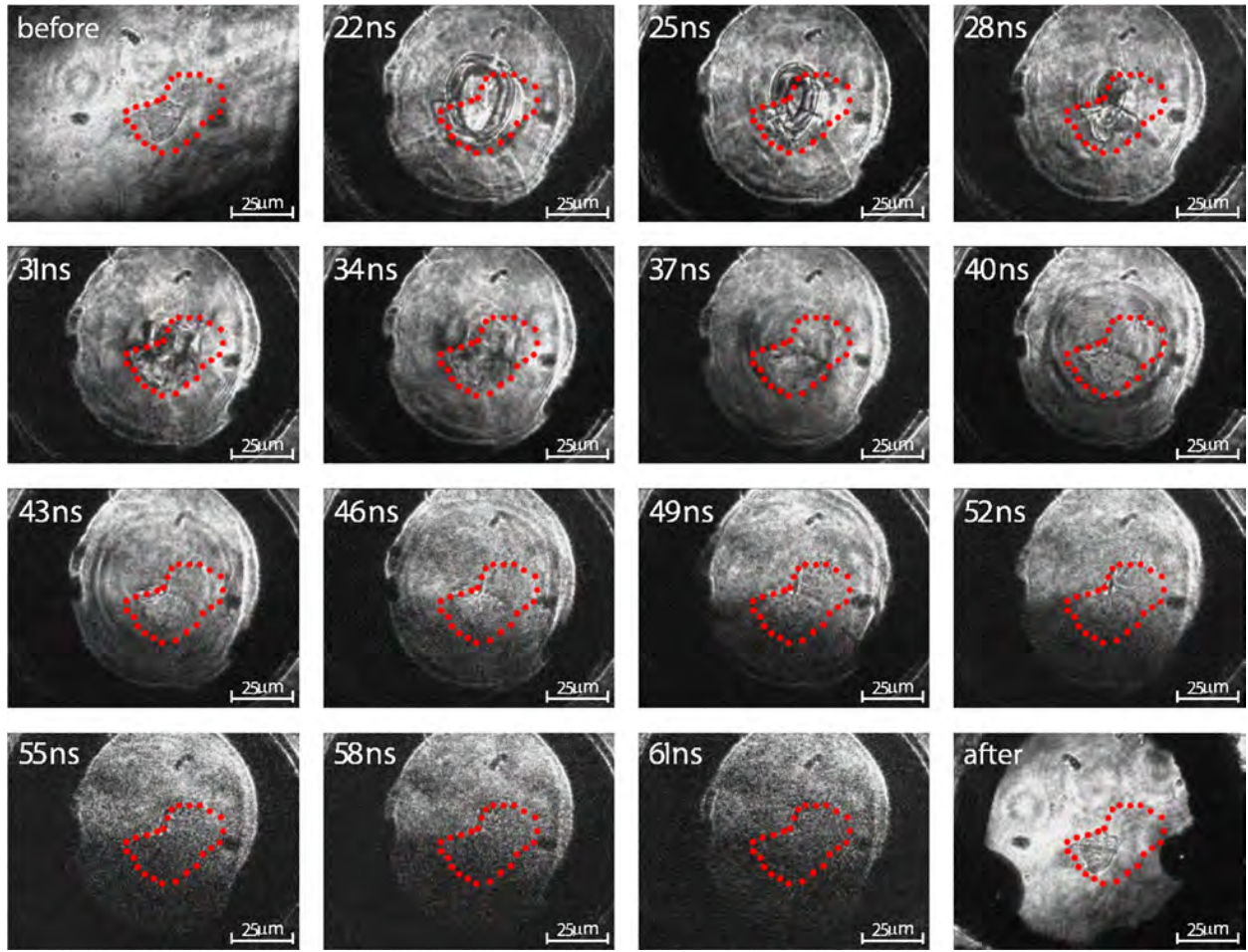

**Figure S4.** *Image sequence showing the shock wave traversing the RDX crystal upon convergence and divergence for a P = 28 GPa shock ($E_{drive}$ = 4.5 mJ). Images from this sequence were collected using shadowgraph imaging, by the fs probe technique described in the manuscript.*

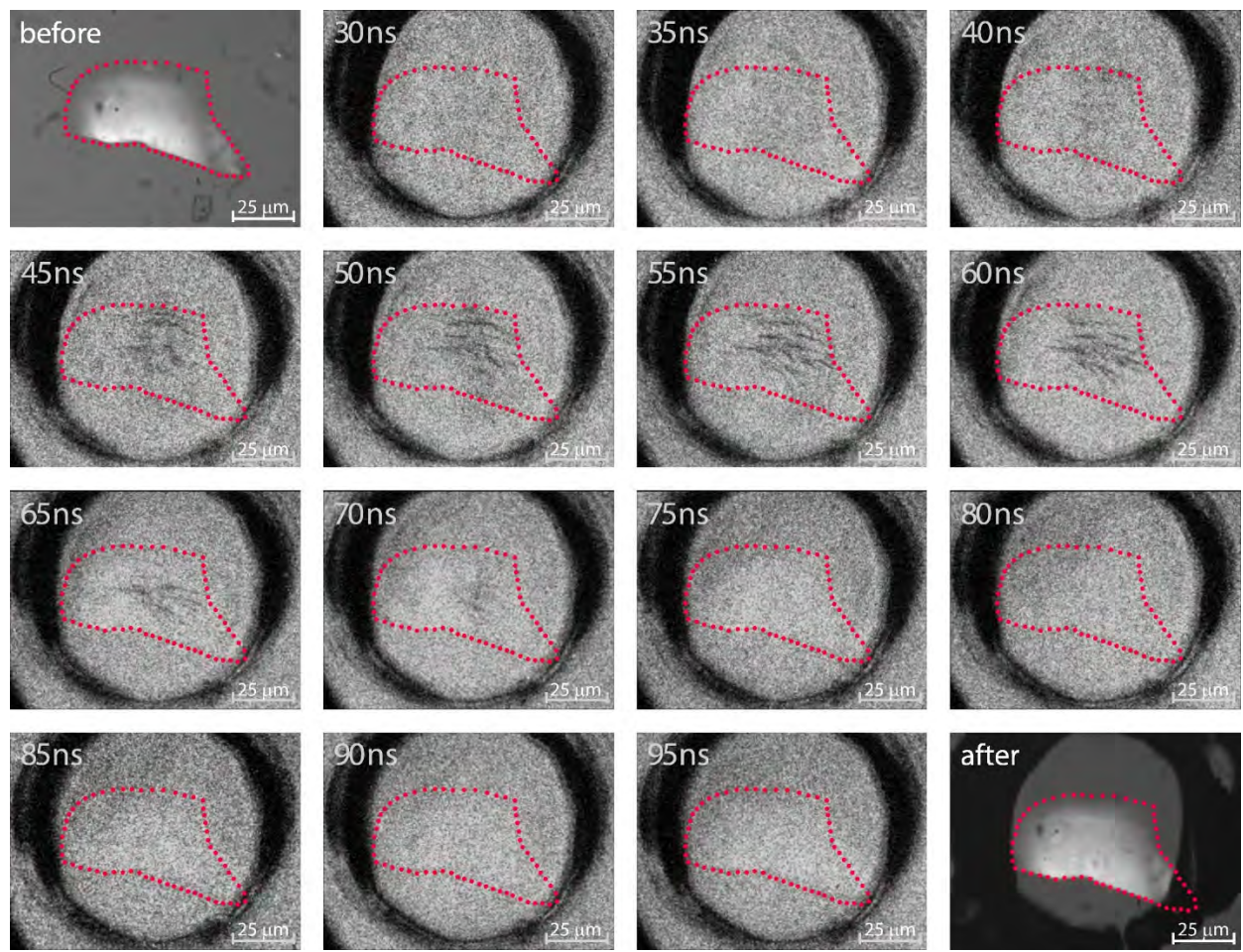

**Figure S5.** *RDX crystal responding to a P = 10 GPa shock wave (E$_{drive}$ = 2.8 mJ). All images were collected with an integration time of 5 ns. All images were collected with non-collimated light through crossed polarizers that were adjusted to maximize the light transmitted through the crystal for the before and after frames and to minimize light transmitted through the crystal during the shock measurement.*

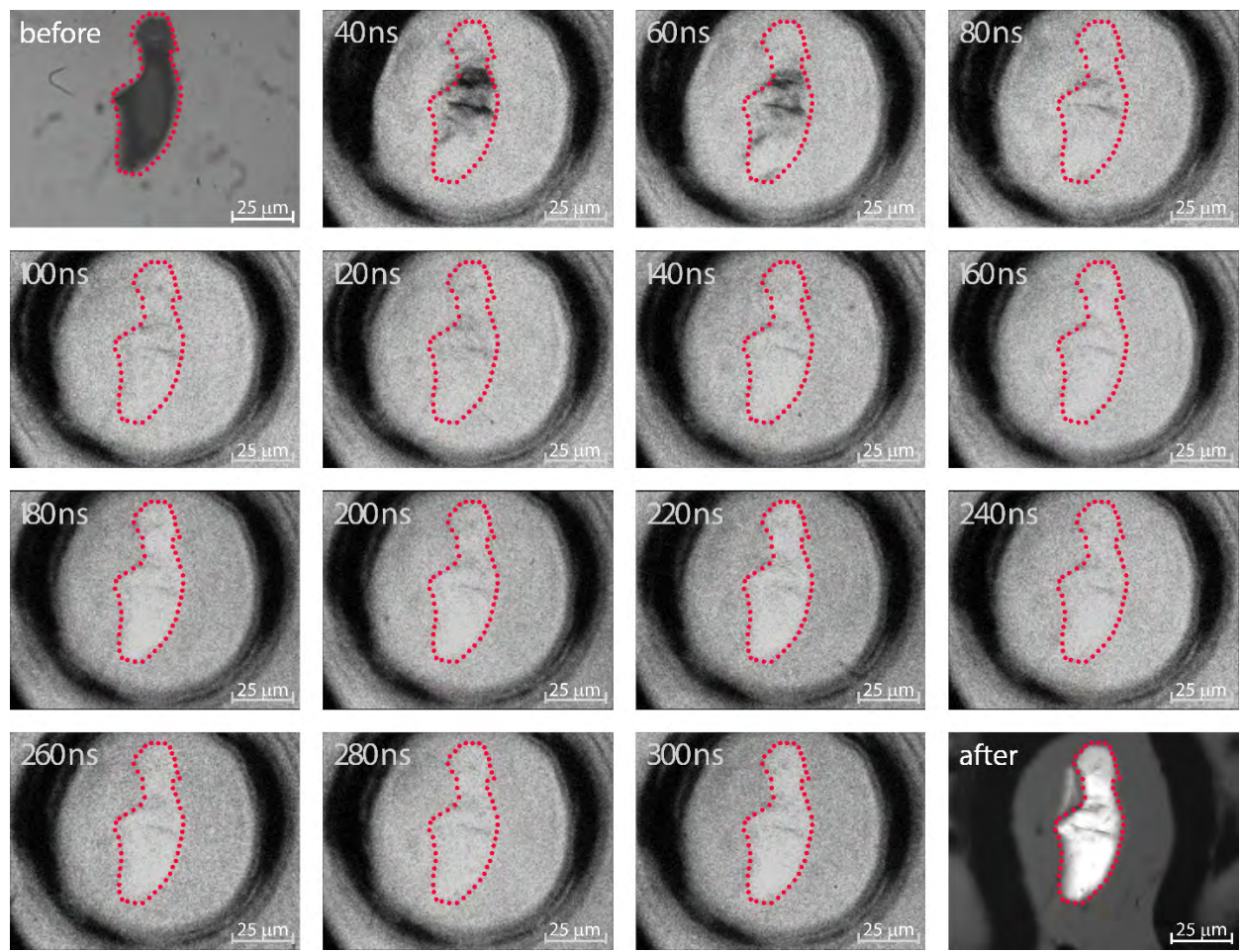

**Figure S6.** *RDX crystal responding to a P = 10 GPa shock wave (E_{drive} = 2.8 mJ). All images were collected with a 5 ns integration time. All images were collected through crossed polarizers that were adjusted to maximize light transmitted through the crystal for the after frame and to minimize light transmitted through the crystal during and before the shock measurement.*

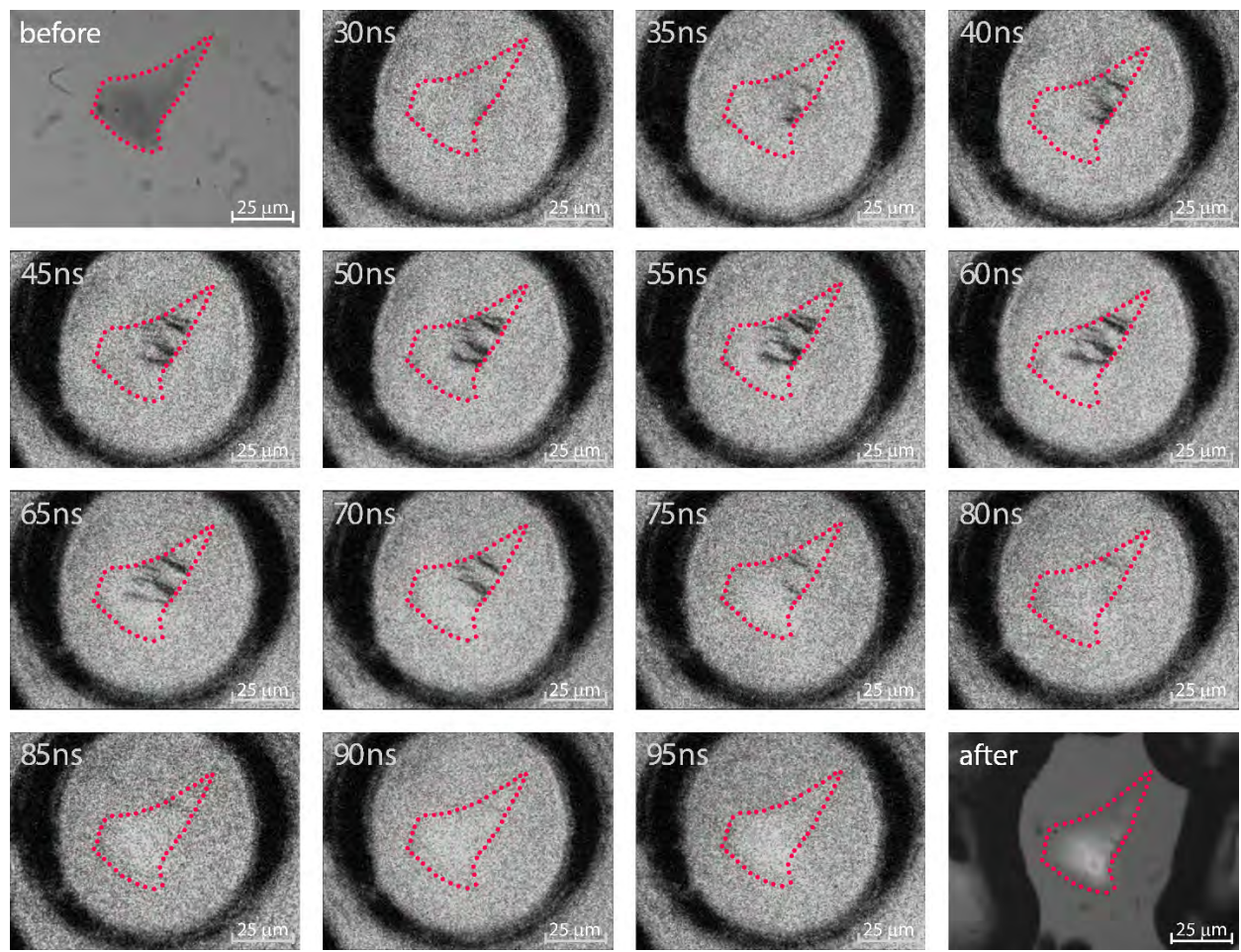

**Figure S7.** *RDX crystal responding to a P = 10 GPa shock wave (E$_{drive}$ = 2.8 mJ). All images were collected with a 5 ns integration time. All images were collected through crossed polarizers that were adjusted to minimize light transmitted through the crystal.*

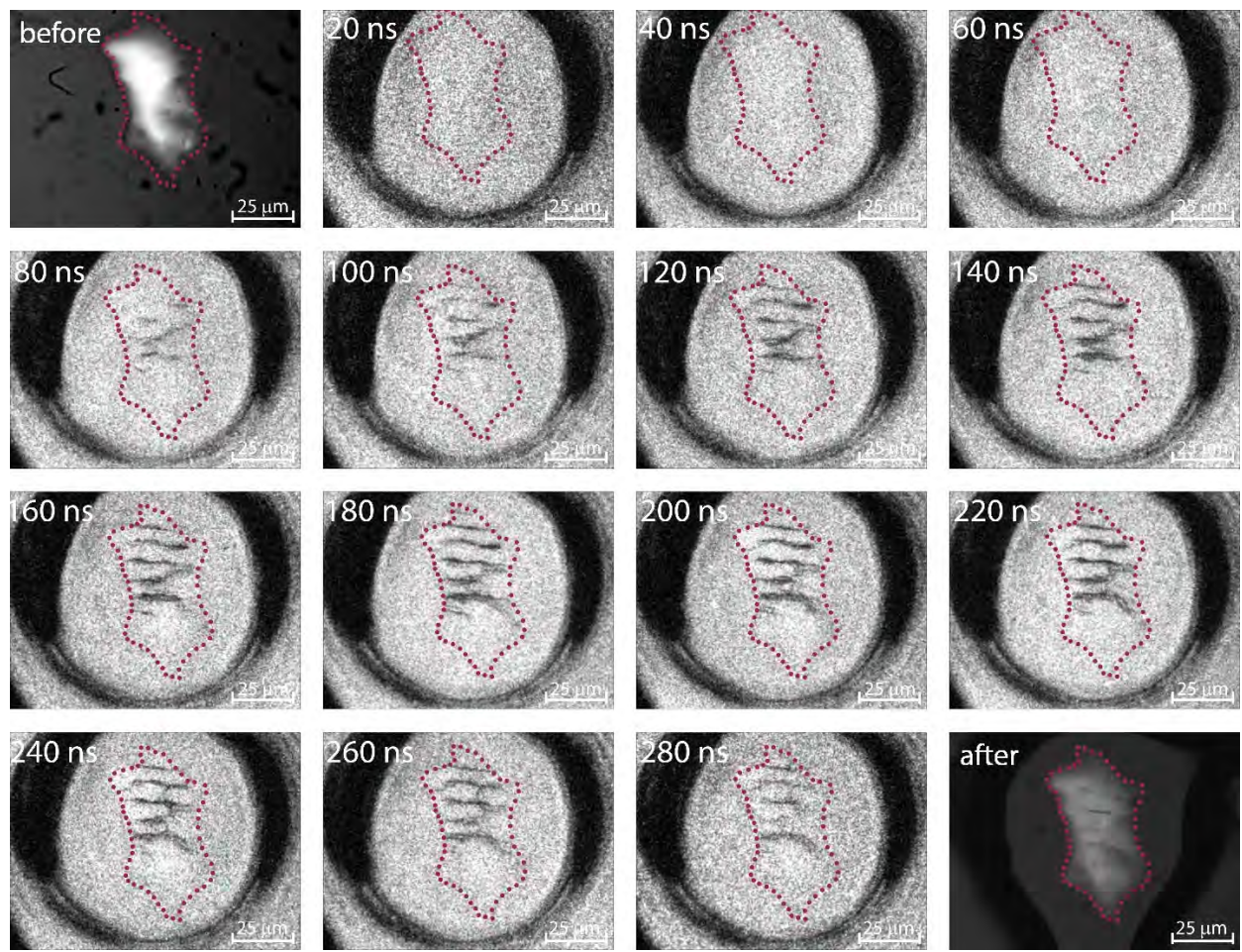

**Figure S8.** *The deformations resulting from a shock with P = 10 GPa traversing an RDX crystal, with E$_{drive}$ = 2.8 mJ. All images were collected with non-collimated light through crossed polarizers that were adjusted to maximize light transmitted through the crystal for the before frame and to minimize light transmitted through the crystal during and after the shock measurement. All dynamic images used a 20 ns integration time. The longer integration time resulted in clearer images than in Figs. S5 and S6, but the nature of the deformations is very similar in all three image sequences.*

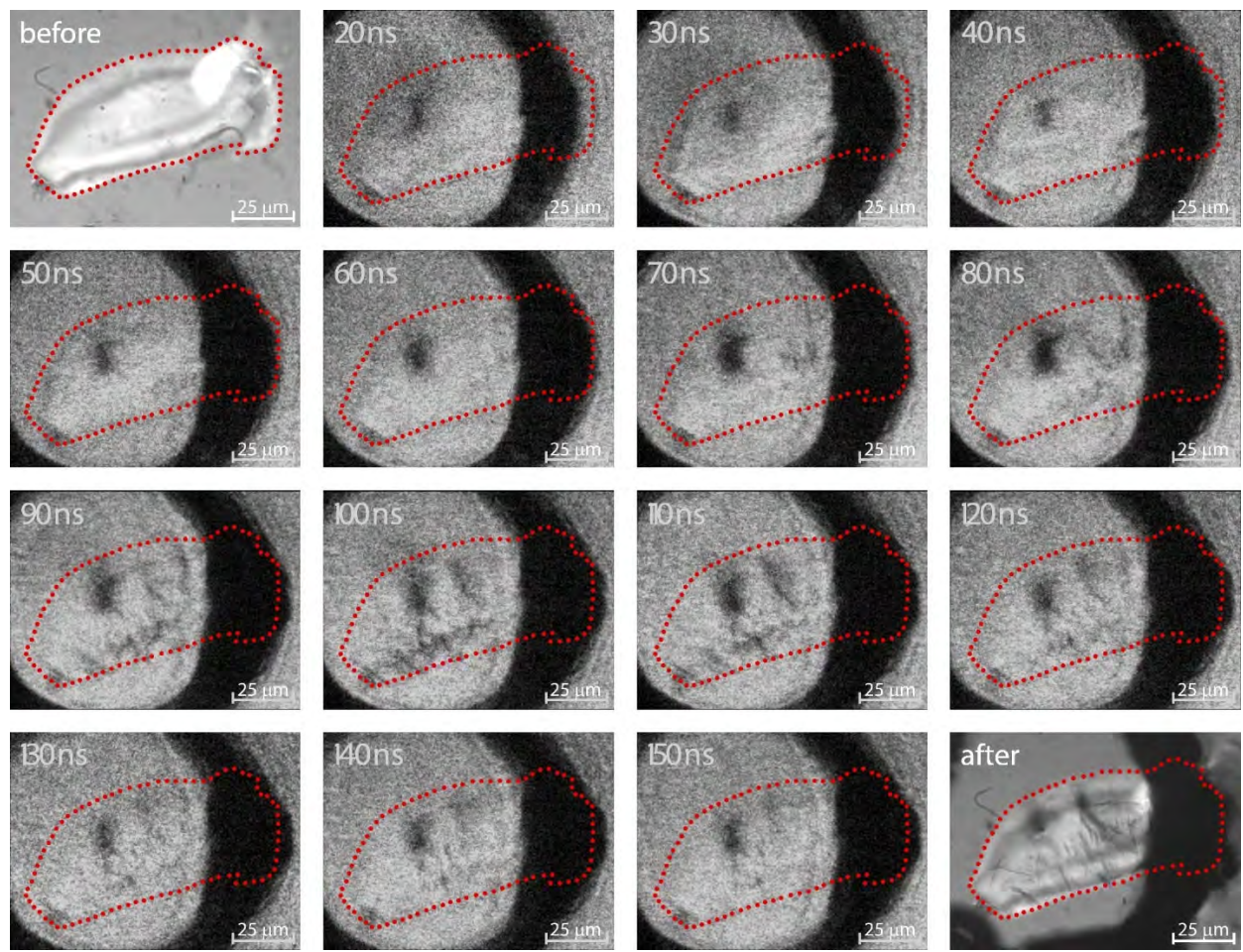

**Figure S9.** *RDX crystal responding to a P = 28 GPa shock wave (E_drive = 4.5 mJ). All images were collected with a 10 ns integration time. All images were collected with non-collimated light through crossed polarizers that were adjusted to maximize light transmitted through the crystal for the before frame and to minimize light transmitted through the crystal during and after the shock measurement.*

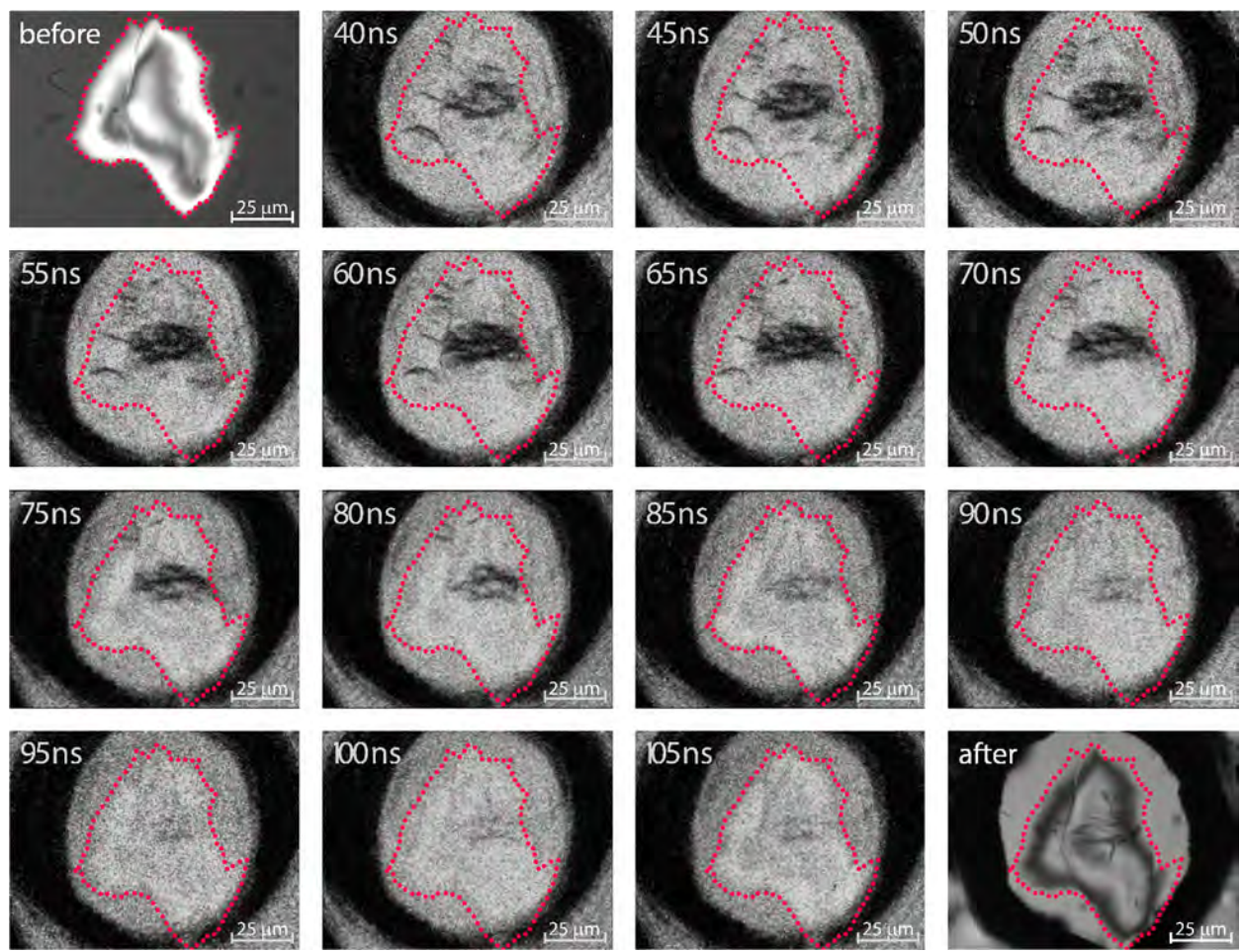

**Figure S10.** *RDX crystal responding to a P = 16 GPa shock wave ($E_{drive}$ = 3.7 mJ). All images were collected with a 5 ns integration time. All images were collected with non-collimated light through crossed polarizers that were adjusted to maximize light transmitted through the crystal for the before frame and to minimize light transmitted through the crystal during and after the shock measurement.*

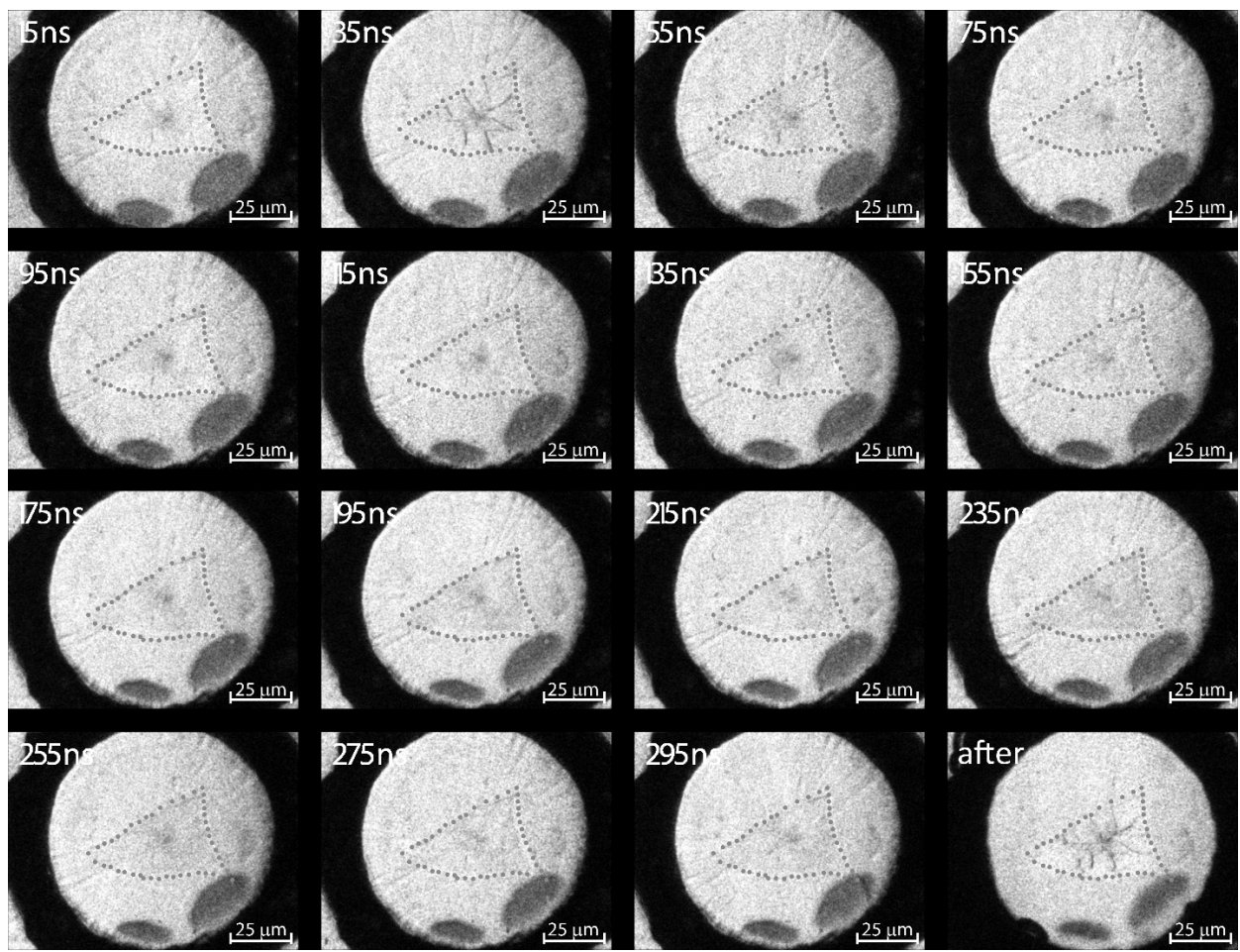

**Figure S11.** *RDX crystal responding to a P = 16 GPa shock wave ($E_{drive}$ = 3.5 mJ). All images were collected using a 10-ns integration time, without crossed polarizers. This image sequence provides a reference for deformations imaged without polarization sensitivity.*